\documentclass[10pt,twocolumn]{article}
\PassOptionsToPackage{hyphens}{url}
\usepackage[letterpaper,margin=0.75in,columnsep=0.25in]{geometry}
\usepackage[T1]{fontenc}\usepackage[utf8]{inputenc}
\usepackage{times}\usepackage{booktabs}\usepackage{microtype}\usepackage{amsmath}
\usepackage{graphicx}\usepackage{array}
\graphicspath{{figures/}}\usepackage{enumitem}\usepackage[hidelinks]{hyperref}
\usepackage{flushend}\usepackage{newunicodechar}\usepackage{textcomp}\usepackage{caption}
\usepackage{xurl}
\usepackage{arxivpreprint}
\newunicodechar{β}{\ensuremath{\beta}}\newunicodechar{≥}{\ensuremath{\geq}}
\newunicodechar{−}{\ensuremath{-}}\newunicodechar{χ}{\ensuremath{\chi}}
\newunicodechar{ρ}{\ensuremath{\rho}}\newunicodechar{²}{\ensuremath{^2}}
\newunicodechar{§}{\S}\newunicodechar{≈}{\ensuremath{\approx}}
\newunicodechar{×}{\ensuremath{\times}}\newunicodechar{Σ}{\ensuremath{\Sigma}}
\newunicodechar{±}{\ensuremath{\pm}}\newunicodechar{≤}{\ensuremath{\leq}}
\newunicodechar{→}{\ensuremath{\rightarrow}}\newunicodechar{α}{\ensuremath{\alpha}}
\newunicodechar{μ}{\ensuremath{\mu}}\newunicodechar{σ}{\ensuremath{\sigma}}
\newunicodechar{ü}{\"u}\newunicodechar{…}{\ldots{}}
\usepackage{stfloats}

\hypersetup{pdftitle={The Machines Are Calling: Measuring Automated and Synthetic Voices in Unwanted Inbound Calls},pdfauthor={Xingyu Shen, Tommy Duong, Muduo Xu, Xiaodong An, Jiaqi Gan, Haoyuan Tang, Jamey Z. Liang, Siyu Zhang, Yan Zhang, Ethan Traister, Simiao Ren},pdfsubject={Measurement of automated and synthetic voices in unwanted inbound calls}}
\arxivstamp{A \textsc{preprint} --- September 14, 2026}
\title{The Machines Are Calling:\\Measuring Automated and Synthetic Voices\\in Unwanted Inbound Calls}
\author{Xingyu Shen, Tommy Duong, Muduo Xu, Xiaodong An, Jiaqi Gan, Haoyuan Tang, Jamey Z. Liang,\\Siyu Zhang, Yan Zhang, Ethan Traister, and Simiao Ren$^{\dagger}$\\[3pt]Scam AI (Reality Inc.)\\[2pt]{\footnotesize $^{\dagger}$Corresponding author: \texttt{benren@scam.ai}}}
\date{September 14, 2026}
\begin{document}\maketitle
{\renewcommand{\thefootnote}{}\footnote{Authorship of this paper does not imply employment by or affiliation with Scam.AI (Reality Inc.); some authors are independent researchers who collaborated on this work.}\addtocounter{footnote}{-1}}

\begin{abstract}In February 2024 the U.S. Federal Communications Commission (FCC) placed AI-generated voices under the Telephone Consumer Protection Act (TCPA)~\cite{fcc2024ruling}. Yet no peer-reviewed measurement says how much unwanted call traffic is placed by a machine, or how much of that machine speech is synthesized rather than played from a recording. We report both with a disclosed pipeline. An interactive voice honeypot (language-model personas on real U.S. numbers, the caller recorded on its own track) recorded 10,987 calls over 66 days. Three instruments read each opening: an audio fingerprint that finds the same recording played on other calls, a commercial synthetic-speech detector on the caller's first ten seconds, and blinded listeners who check what it flags. Of the 7,233 greeted calls we analyze, \textbf{13.8\% open with a recording we also heard on another call, and 13.1\% with fresh audio the detector labels synthetic}. A further 9.9\% open with a caller who never spoke after our greeting, 54.2\% with fresh audio the detector labels human, and 9.0\% could not be scored. Machine-voiced openings are therefore at least 26.9\%, a further tenth of calls are silent connections we read as machine-placed, and replays of a recording make up 45\% of the detector's own rate (29.3\% of 6,192 scored openings). The same waveform played on two calls lands on opposite sides of the detector's threshold 13.6\% of the time, and eleven listeners confirm 54.4\% of what it flags. Synthetic openings concentrate in lead-generation spam (33.8\%), not fraud (21.1\%); 0.44\% disclose automation. Prevalence tracks how long a bait number has circulated (59\% against 19\% in the same weeks): seeding history, not calendar time, explains the trend. Campaigns outlast their numbers: one recorded compliance notice opens calls in six campaigns, and one synthetic voice serves nine.\end{abstract}
\keywords{telephony abuse $\cdot$ voice honeypot $\cdot$ synthetic speech detection $\cdot$ robocalls $\cdot$ measurement}
\section{Introduction}

Telephone fraud has a new input. Putting a natural-sounding voice on a call once meant hiring a voice actor and a studio. It is now a metered call to a cloud speech service, and both journalists and vendors report its arrival in scam calling. Policy has moved accordingly: the Federal Communications Commission's (FCC's) February 2024 declaratory ruling~\cite{fcc2024ruling} brought AI-generated voices within the Telephone Consumer Protection Act (TCPA), and carriers market deepfake detection as a consumer feature.

What is missing underneath all of this is a number, and it is two numbers. How much of the unwanted traffic reaching consumers is placed by a machine rather than a live person, and of that, how much is a synthesizer speaking versus a recording being played? The answers determine whether AI voice is a marginal novelty or a dominant delivery mechanism, and whether detection belongs in the network or at the handset. They also say whether a rule written in 2024 addressed a real distribution of harm or an anticipated one.

By \emph{unwanted} we mean what the recipient means: a call they did not want when it came, whatever consent was collected earlier. A box ticked once on a web form does not make every later call wanted, and the record behind it is resold down a chain the consumer never agreed to (§4.4). Our recipient was a fabricated consumer record, so no call to it served a real prospect; many callers nonetheless held a consent box our seeding agent had checked, which is why the consent question is only half observable (§3.1, §5.1).

A second question sits behind the first. The 2024 ruling attached the TCPA's existing duties to synthetic voice: consent, identification of the entity responsible, and opt-out. The Commission has since proposed, but not adopted, a rule requiring callers to say that a call is AI-generated. No published work reports how often a caller already says that the voice on the line is a machine. Our corpus supplies that baseline, and it is close to zero.

Measuring either requires the caller's voice by itself. Our honeypot records it that way by construction. Each conversational large language model (LLM) persona answers on its own telephone number, and the caller's audio is written to a channel separate from the persona's synthetic speech, so a detector scores the caller and nothing else. Prior honeypots kept either no audio or a mixed line, on which any score would blend the caller with the honeypot's own text-to-speech. §3 describes the scoring pipeline.

Picking up the phone also removes a bias that passive honeypots cannot escape. A sophisticated campaign may decide that a silent line is not a live person. A honeypot that greets and then goes quiet therefore misses the interactive, live-agent end of the traffic, and any synthetic fraction it computes is biased toward prerecorded blasts. Ours greets and listens, and it records every call whether or not the caller ever spoke, so the calls that never became a conversation are counted rather than filtered away (§4.1).

Whether an automated caller is playing a file or speaking afresh is a different question from whether it reached us, and it is one that no synthesis detector answers. We answer it with an audio fingerprint: the same recording played on two calls leaves the same waveform, whatever its voice, and a synthesizer rendering a script afresh does not (§3.5). That instrument needs no training data and makes no claim about voices; it is the part of the measurement that depends on no vendor. It has a threshold of its own, which §4.1 reports at both of its settings.

Our primary instrument is a closed commercial detector. We cannot inspect its training data, and we apply it to narrowband telephone audio, where published detectors lose much of their accuracy once they leave the conditions they were trained on (Appendix A). We therefore treat it as a screening tool and measure its error rate ourselves: blinded listeners judge the same ten-second clips it scored. Of the 1,816 flagged calls, 878 carry a blinded judgment, and listeners confirm 54.4\% of them (§4.2). The calls the detector did not flag, which would turn that precision into a prevalence, carry only 23 judgments, so this paper reports a precision, not a prevalence (§5.5). Every detector-labeled rate below is therefore paired with the share of it that listeners confirmed, and with the detector-free decomposition of §4.1.

\subsection{Research questions}

We ask four questions of this corpus.

\textbf{RQ1. How much inbound unwanted call traffic is machine-voiced, and how much of that is a recording rather than a synthesizer?} Of the calls our persona greeted, 26.9\% open with a machine voice on the audio evidence alone: 13.8\% with a recording we also heard on another call, and 13.1\% with fresh audio a commercial detector labels synthetic. A further 9.9\% are silent connections on which nobody spoke after our greeting, and 9.0\% could not be scored (§4.1). The split moves with the fingerprint's own threshold: at its stricter setting the recording share is 9.2\% and the synthetic share 16.7\%, while the machine-voiced total moves by a point (§4.1).

The detector by itself labels 29.3\% of the 6,192 scored openings synthetic, and 45\% of those are replays. Blinded listeners confirm 54.4\% of what it flags, which puts human-confirmed synthetic openings at 15.9\% of scored calls under a transfer assumption we cannot test, or 15.6\% once the labeled clips are reweighted to the census's score mix (§4.2). To our knowledge these are the first such figures whose thresholds, score distributions, aggregation rules and engagement model are disclosed well enough to be argued with. The one prior public number is an industry report that discloses none of these.

\textbf{RQ2. Where in the traffic do machine-voiced calls sit?} In the high-volume, low-harm part of it. Calls the detector labels synthetic are more common among lead-generation sales calls (33.8\%) than among calls our classifier labels fraud (21.1\%). They are far more common among calls that end with the caller handing us to a second person on their side, typically a sales closer (§4.3).

We expected four further differences and found none that survives scrutiny (§4.6). AI-labeled calls (those the detector flags as synthetic) do not convert fewer targets once length is accounted for. They ask for credentials less often only because they end sooner. They are not an automated front end that hands to a live closer. And their conversational features do not split them into prerecorded blasts and conversational agents; the audio fingerprint makes that split for them. AI-labeled calls are shorter, but mostly because they come from different kinds of phone numbers (toll-free rather than local or geographic) than human-voiced calls do. They do answer beside the point: an automated caller ignores one of our questions in three, a human-labeled one in four (§4.6).

\textbf{RQ3. How often do automated callers disclose that they are automated?} On 0.44\% of calls whose opening we label synthetic; 99.6\% never say so, though 17\% announce that the call is recorded (§4.5). No federal rule currently requires disclosure, so the 0.44\% is a baseline for a duty that does not yet exist, not a compliance rate. Counting it at all requires separating TCPA autodialer-consent boilerplate, a request for permission to autodial on future calls, from an actual admission that the present call is automated.

\textbf{RQ4. What part of a calling operation actually persists?} The words and the assets; the number is disposable, but no more so inside a campaign than in this corpus generally. Grouping calls that open with near-identical wording gives us campaigns. A campaign keeps reading the same opening line for weeks while the number it dials from changes on nearly every call; in the sharpest case, 68 calls came from 68 different numbers over 31 days. That rate of churn, however, is what the corpus's number supply gives any set of calls (a permutation null, §4.7). What campaigns specifically add is a minority that dials from one number for weeks. Beneath the scripts sit shared assets: one recording of a compliance notice opens calls in six different campaigns, and single synthetic voices serve as many as nine (§4.7).

The distinction matters because blocklists, caller reputation and carrier analytics all identify a caller \emph{by its number}. For most operations they therefore track the part that is thrown away, the number. They can act only after enough people have been called to give that number a bad name, by which time it has been abandoned. The opening line, which is what stays constant, is already available to anyone transcribing the call (§5.2).

One methodological finding cuts across the four questions. A honeypot's seeding history is when each of its bait numbers was planted in the lead-generation forms that draw the calls. That history manufactures an apparent ecosystem trend for anyone who does not control for it (§4.4).

We also describe the blinded protocol for validating synthetic-speech detection on real 8 kHz telephony that produced the 54.4\% estimate (§3.4). And we add a detector check that needs no listener: the same waveform played on two calls crosses the detector's threshold on roughly one pair in seven (§4.2).

\section{Related Work}

\textbf{Telephony abuse measurement.} The empirical study of unwanted calling matured around large-scale honeypots, but its instruments were built to characterize volume, spoofing and campaign structure, not voice provenance~\cite{tu2016sok,sahin2017sok}. Phoneypot~\cite{gupta2015phoneypot} logged 1.3M calls across 39,696 numbers but answered none of them, rejecting every call with a busy tone and retaining only the calling number, called number and timestamp. Number blacklists built from complaint and honeypot data block more than 55\% of unwanted calls at a 0.01\% false-positive rate~\cite{pandit2018blacklists}.

\emph{Who's Calling?}~\cite{prasad2020whoscalling} ran a honeypot of up to 66,606 lines for eleven months, mechanically answering a subset of calls with a fixed greeting (a non-interactive honeypot, in its authors' later terminology). It clustered the recorded calls into campaigns by audio fingerprint, and stated the construct limit precisely: ``what we are characterizing as campaigns is audio, not operators.'' That limit is central to our argument. A prerecorded human voice replayed across thousands of dialed numbers clusters exactly as a text-to-speech rendering would, so audio similarity collapses the distinction we set out to measure. SnorCall~\cite{prasad2023snorcall} applied weak supervision to 232,723 transcripts to characterize what robocalls say; its only mention of speech synthesis is incidental, as poor-quality text-to-speech (TTS) output corrupting the extraction of dollar amounts.

\emph{Characterizing Robocalls with Multiple Vantage Points}~\cite{prasad2025vantage} drew on five vantage points: two voice honeypots, Federal Trade Commission (FTC) complaints, enforcement-action recordings and an Internal Revenue Service (IRS) data feed. It compared audio and transcripts from roughly 3M voice calls. It contributed a technique for classifying a campaign as interactive or static. It also observed that an interactive honeypot captured more analyzable audio than a non-interactive one (66.8\% versus 42\% of calls with more than 10\% audio), and nearly ten times as many callback numbers. It comes closest to our question without answering it: interactivity is a campaign-behavior axis, and an interactive campaign may be a live agent, an interactive voice response (IVR) menu, or an LLM driving TTS. That paper anticipates that ``easy-to-use generative-AI based audio services will likely accelerate this change'' and calls for ``deployment of interactive honeypots to study `smarter' robocall campaigns.'' Voice provenance is orthogonal to interactivity, and to our knowledge unmeasured. This paper answers that call.

\textbf{Scambaiting and AI-voice capability.} A parallel literature deploys agents to engage scammers, but measures detection in the opposite direction. Lenny~\cite{sahin2017lenny}, the field's only published human-versus-bot measurement, is a scambaiting bot whose study asks whether the human scammer notices the \emph{defender's} automation. It found recognition in only 11 of the 200 transcribed calls (about 5\%), sampled from 487 public recordings. Later work moved to simulation and offline analysis~\cite{basta2025botwars,wood2023scambaiting}.

Capability research establishes that LLM agents can execute scam scripts~\cite{fang2024llmagents} and that surveyed targets say they would or might comply at rates up to 36\% in the most effective scam category~\cite{heiding2026compliance}. Both are laboratory measurements, and neither reports field incidence; a field experiment on live calls did show that people answer and engage with scam calls~\cite{tu2019users}. The clearest evidence of the gap is RoboKA~\cite{roboka2026}. It synthesized its own corpus of roughly 1,200 unwanted calls across four TTS systems because no public labeled real-world corpus was available to train on, and kept 1,378 real recordings only for out-of-distribution evaluation.

\textbf{The one competing prevalence estimate.} Hiya operated a honeypot across more than 100,000 owned numbers in early 2025 and reported that approximately 25\% of the 8,455 calls with sufficient speech contained AI-generated audio~\cite{hiya2025}. The report is unrefereed. It uses Hiya's own AI Voice Detection product and discloses none of its threshold, score distribution, aggregation rule or engagement model. Our detector-labeled 29.3\% lies in the same range; the contribution here is not priority but auditability. Pindrop has reported rapid growth in detected deepfake traffic to enterprise contact centers~\cite{pindrop2025}, a different denominator from outbound consumer campaigns. The regulatory environment has moved ahead of all of it: the FCC's February 2024 ruling~\cite{fcc2024ruling} brought AI-generated voices under the TCPA with no published baseline for how common such voices already were.

\section{Data and Method}

\subsection{The corpus}

The honeypot and its data are documented by a companion data descriptor~\cite{scamai2026corpus} (in preparation) and summarized in a prior analysis of 10,211 calls from this collection~\cite{traister2026anatomy} (that analysis counts logged sessions; this paper counts recordings, §4). That analysis characterizes call volume, scam taxonomy and conversational structure without classifying the caller's voice. The corpus is not a contribution of this paper. We describe it only as far as reading the results requires, and refer the reader to~\cite{traister2026anatomy} for collection and labeling and to the descriptor~\cite{scamai2026corpus}, once released, for de-identification. Appendix B states this paper's own screening and release commitments.

In brief: conversational LLM personas, each on its own real U.S. telephone number, answer inbound calls and converse with the caller. The numbers were seeded into U.S. lead-generation funnels, web forms that collect a consumer's contact details for resale to marketers, using a fabricated identity; that is what generates the traffic (Appendix B). Where a form presented a consent-to-contact checkbox, the seeding agent checked it. Collection ran from 28 May to 21 July 2026; the recordings analyzed in §4.1 also include ten days of testing before it and one day after.

Every call is recorded with the caller's audio and the persona's audio on separate channels, so the caller's speech can be scored without our own text-to-speech contaminating the signal. Separate channels are the precondition for the measurement (§1).

The seeding history bears directly on §4.4. For most of the period a single bait number received all inbound traffic; on 1 July 2026 ten further numbers were seeded, one per persona, taking the fleet from one receiving number to eleven. A twelfth, low-volume line received nine scored calls and is omitted from the exposure-age analysis. The ten entered with zero exposure age (days since a number's first inbound call) while the original had circulated since May. That staggered seeding is what separates exposure age from calendar time, and why the pooled prevalence below must be read under this seeding schedule rather than as a transportable constant.

Every inbound call is recorded whether or not it became a conversation, and those recordings are the denominator of §4.1: 10,987 calls with a caller track over 66 days, test-line calls removed. The two-turn corpus of our prior analyses is a stratum inside that. It keeps real inbound traffic, excluding test lines, sessions that were not genuine calls, and calls with fewer than two logged turns, which yields 6,619 calls, of which 6,192 carry scorable caller speech. The remaining 427 could not be scored: 5 had no speech energy at all, 110 had an onset but no clip could be cut, and 312 were cut and submitted but returned no scored utterance. Detector-labeled rates from §4.2 onward are over those 6,192 unless noted. The two-turn filter drops 1,845 calls on normal days; §4.1 shows they are mostly silent connections, not prerecorded blasts.

Two facts about the persona fix how a silent recording is read. The persona always speaks first, with a fixed ``Hello?'' in the ten fleet personas and a generated one in the original. It never re-prompts, times out or hangs up: a call ends only when the caller's side disconnects. A caller who is silent on the recording therefore heard our greeting and chose to say nothing for as long as they stayed on the line. The logged turn count includes that greeting, so the two-turn filter means ``our greeting plus at least one transcribed item,'' and a caller who never produced transcribable speech falls outside the corpus.

On eleven of the 66 days a system fault left the persona silent after answering, so more than half of that day's callers met dead air. Those 2,711 calls follow a stated rule. A call on which neither side spoke is dropped, because a person may have been waiting for us to speak. A call on which the caller played audio heard on another call, or read a script seen on another call, is counted as a machine even without our greeting. That is 188 of the 1,326 outage-day calls with caller speech, and 71 of the 454 normal-day calls that got no greeting but on which the caller spoke. The rest, 1,138 outage-day calls and 383 normal-day calls, are undetermined (Appendix A).

\subsection{Isolating and windowing the caller's speech}

Every step operates on the caller leg alone. The opening seconds of an inbound call are dominated by ringback, hold-queue noise and pickup silence, so a fixed leading window would mostly measure non-speech. We locate the onset of caller speech with an energy-based detector whose threshold is set relative to each file's own noise floor rather than an absolute level. The onset detector operates on 30 ms frames with a 10 ms hop, estimating the noise floor as the 10th percentile of frame energies. An onset is declared where frame energy exceeds that floor by at least 10 dB continuously for 250 ms; we retain 200 ms of pre-roll. From that onset we cut a fixed ten-second window.

Ten seconds is chosen from the antispoofing literature rather than by default. Detector error climbs steeply below four seconds of speech: on the ASVspoof 2021 logical-access track, AASIST2~\cite{zhang2024aasist2} reports equal-error rates of 11.7\% at one second, 3.37\% at two and 1.76\% at four. Ten seconds also matches the ASVspoof 5 operating point~\cite{wang2024asvspoof5}.

Onset alignment closes a documented failure mode. Müller et al.~\cite{muller2021silence} show that leading-silence duration alone classifies bona fide versus spoofed audio at roughly 85\% accuracy on ASVspoof-family corpora~\cite{todisco2019asvspoof,asvspoof2021}. A detector trained there has therefore partly learned that long leading silence implies a live speaker. A telephone call opens with exactly that silence. We cannot verify the mitigation against a closed synthesis detector.

\subsection{Detector, aggregation and threshold}

Each clip goes to a commercial conversation-intelligence service, reached through its application programming interface (API). We do not name the vendor: the product is a deployed synthetic-speech detector with a voice-phishing preset, used exactly as shipped, with no tuning or adaptation to this corpus, and nothing in this paper depends on which one it is. The service segments each clip into per-utterance spans and returns a \texttt{deepfake\_score} in $[0,1]$ per span together with a speaker label, an accent classification and a transcript. Because the caller leg carries a single speaker, attribution is unambiguous.

We aggregate to one per-call score by taking the maximum across utterances (df$_{max}$ in the figures), reasoning that a single clearly synthetic utterance is sufficient evidence the caller is not live. We report the mean as a sensitivity analysis. The maximum is a logical-OR rule: one utterance over the line flags the whole call. Its false-positive rate grows with utterance count, and utterance count is not uniform across strata, so §4.2 reports the mean-aggregated rate alongside the maximum.

We label a call synthetic when its maximum per-utterance score reaches 0.85. We do \textbf{not} claim this is the valley of the score distribution. The trough lies near 0.45--0.50: in 0.05-wide bins, 64 calls fall in the 0.45--0.50 bin against 106 and 113 in the two bins adjacent to our threshold. A threshold placed at the trough would yield 39.6\% rather than 29.3\%. Our operating point sits on the rising shoulder of the upper mode and is therefore the conservative end of the defensible range. Because a point estimate is sensitive to this choice, §4.2 sweeps it.

Transcripts used for the text-derived labels in §4 come from the persona's own real-time automatic speech recognition (ASR), written to the message log during the call, and not from the detector API. The text and audio instruments are therefore separate.

\subsection{Validation protocol and estimator}

\textbf{Who listens, and to what.} To measure the detector's error rate on our own data rather than from published benchmarks, eleven listeners internal to the authors' organization label ten-second clips through a blinded interface at \texttt{canuspotai.com}. They were given no training, screening or attention checks. Clips are cut by the same code path, from the same onset-aligned ten-second window, that the detector scored. Before serving, spans containing personal information are silenced (about 8\% of clips carry any mask) and the clip is encoded as a 64 kbps mono MP3. Listener and instrument therefore hear the same window, though not byte-identical audio.

\textbf{Strata and pool.} The design is stratified: every call flagged at the operating threshold (1,816, a census), a random sample of unflagged calls to estimate the false-negative rate, and gold-control clips of known provenance. All three are served through the same priority tiers, with order randomized within each tier so that clip order cannot leak the stratum (the tiers are described below). As run, the flagged stratum was served from a wider pool selected at a maximum per-utterance score of 0.50 or above, so listeners also heard clips scoring between 0.50 and 0.85 (§4.2 uses them).

The served pool is not the flagged census. Of the 1,816 flagged calls, 1,348 entered it and 878 carry at least one decisive judgment (human or synthetic, as opposed to unsure). A clip was dropped where masking personal information would have silenced more than a third of the window (21 clips). Clips with fewer than twelve transcribed words were withheld as too thin to judge, and the number of active clips from any one near-identical opening script was capped. The labeled clips are therefore not a random subset of the flagged census.

They are longer (median 121 s against 90 s for unlabeled flagged calls) and higher-scoring (87\% against 78\% at a maximum score of 0.95 or more). Repeated scripts also score higher than singletons within the pool. Precision is therefore bootstrapped over labeled clips rather than treated as a census value. Because confirmation rises with score (§4.2), the labeled clips' 54.4\% slightly overstates precision on the census; reweighting them to the census's score mix gives 53.2\% (§4.2), and any transfer to all 1,816 remains an assumption.

\textbf{The interface.} It presents one clip at a time and asks whether the voice is human or synthetic, with \emph{unsure} as the third option. It plays the full ten-second window the detector scored, so the listener is at least as well informed as the instrument under evaluation. It displays a two-second minimum before a judgment, which is advisory rather than enforced. And it logs decision latency and the milliseconds actually consumed, counting a passage that is replayed three times only once. It never reveals whether a judgment agreed with the detector, since a running score would teach listeners the answer and destroy the measurement. Listeners never see the detector's score, the stratum, the campaign, the transcript, or another listener's judgment.

\textbf{What the pool can measure.} Because the deployed pool is drawn from detector-flagged audio, it carries no human-truth negative class. It therefore yields the \emph{precision} of the flag, the share of flagged clips a listener also calls synthetic, and cannot yield recall, listener accuracy, or a false-positive rate. We report precision only, and never describe a listener as correct or incorrect. In a pool where every clip is detector-flagged by construction, a per-listener ``accuracy'' against the flag would measure only how often that listener pressed one button, so we do not compute one.

\textbf{Three properties that shape the estimator.} First, clips are served in priority tiers, unlabeled first, then those that drew an \emph{unsure} or a split, so ambiguous clips accumulate labels fastest. We therefore compute the synthetic fraction \textbf{per clip and then average over clips}, bootstrapping over clips. We also report no agreement coefficient computed within a fixed label-count group, since clips still carrying exactly two labels are the survivors of a tier that re-serves disagreements. Second, listeners vary: mean pairwise Cohen's $\kappa$ across the 23 listener pairs with at least 25 clips in common is only \textbf{+0.267} (median +0.357). We report the all-listener estimate as primary and pre-specify one exclusion as a sensitivity analysis: a listener who agrees with the majority of their peers less than half the time, which one of eleven does. Third, the interface records the audio actually consumed, so we report listening-duration sensitivity alongside it.

\textbf{Round two.} A second round addresses the stimulus rather than the listeners. Across the 2,382 screened clips the ten-second window contains a median of only 4.1 seconds of actual speech (49\% under four seconds), since call openings are dominated by ringback, hold noise and pauses. The clips listeners actually judged in round one passed a twelve-word transcript gate and carry a median of 5.8 seconds, with 18\% under four. Round two re-serves the 330 clips whose round-one judgment was split or drew an \emph{unsure}, this time as roughly 20 seconds of speech with silence removed. A voice-activity-detection (VAD) pass extracts the speech segments. Pauses under 400 ms are kept as original audio, and longer gaps are replaced by a 180 ms pause. On the 13\% of clips whose speech would otherwise exceed a 26-second budget, the pause is shortened, and the amount is recorded per clip. Encoding is unchanged, so file properties still separate nothing.

Round two also carries 100 control clips on which round one's two decisive judgments agreed (62 both human, 38 both synthetic). The interface's consensus rule is two agreeing votes with no dissent; the controls are drawn at random and are indistinguishable to the listener. Without that arm, better agreement on contested clips could not be told apart from every listener simply shifting toward ``synthetic.'' The two rounds never run simultaneously, because a twenty-second clip is distinguishable from a ten-second one in the interface. And because the detector scored the ten-second window, a listener hearing twenty seconds is no longer judging the same evidence as the instrument. Round two therefore measures how much of the disagreement is task difficulty rather than detector error, and its judgments are not pooled with the precision estimate above.

\textbf{A floor on false negatives that needs no listener.} Where a caller states in its own words that its call is automated and the detector scores it below threshold, the label is wrong by the caller's own account. Four such calls are visible here, scoring 0.089, 0.107, 0.207 and 0.404. The IVR greeting ``Welcome to [government program]'s automated customer survey'' opens three calls scored 0.94, 0.21 and 0.09, so the same script sits once in Table~\ref{tab:disclosures} (§4.5) and twice below threshold. These calls are a hand-verified floor on the false-negative rate, independent of the listening study, and the reason we report disclosure only within the flagged population (§4.5).

The same audio content can also fall on both sides of the threshold: the banner ``AI is taking notes'' opens 15 calls scored 0.97--0.98 on fourteen of them and 0.14 on the fifteenth. Identical scripted content drawing near-opposite scores with high confidence at both ends indicates that the detector's errors are not confined to the decision boundary.

\textbf{Estimator.} Because the flagged stratum is a census while the unflagged stratum is sampled, pooling naively would overweight the flagged population. Let $N_f$ be the flagged census and $\hat{p}_f$ the fraction human review confirms synthetic; let $N_u$ be the unflagged population, $n_u$ the sample drawn from it, and $\hat{p}_u$ the fraction of that sample confirmed synthetic. Corpus prevalence is

\begin{equation}\hat{P} \,=\, \frac{N_f \hat{p}_f \,+\, N_u \hat{p}_u}{N_f + N_u}.\end{equation}

Under the design $\hat{p}_f$ is a census value and all sampling uncertainty sits in $\hat{p}_u$. We therefore bootstrap the unflagged sample alone, clustered on originating number, and report $\hat{P}$ under both extreme treatments of ``unsure'' as bounds. As run, $\hat{p}_f$ rests on 882 labeled clips and carries its own clip-bootstrap interval (§4.2). Census or sample, $\hat{p}_f$ carries listener error, so the resulting interval describes prevalence as this protocol labels this corpus, not true prevalence.

\subsection{Three instruments that need no listener}

\textbf{Replay census.} Whether a caller's opening was played from a file or produced afresh is a question about waveforms, not voices. For every recorded call we cut a fifteen-second window from the caller track at the onset of caller speech, using the detector's onset where it scored the call and an energy onset otherwise. From each window we compute 20 mel-frequency cepstral coefficients every 10 ms and normalize each coefficient within the window. Then we compare every window against every other.

A fast Fourier transform cross-correlation over lags of up to five seconds nominates the sixteen best-matching windows for each call. Each nominated pair is then confirmed by the cosine similarity of the two feature sequences at their best alignment, refined to 10 ms. Unrelated speech scores about 0.1 under this measure, the same script rendered afresh by the same voice 0.2--0.5, and the same file played twice 0.85--0.98. We call a pair the same recording at 0.70 and the same waveform at 0.85, and take connected components over confirmed pairs as replay groups.

The replay row of Table~\ref{tab:decomp} uses the 0.70 groups; the band between 0.70 and 0.85 is not certified by that calibration, which is why §4.1 also reports the 0.85 census. Windows with no usable audio (3,173 with no speech energy in the window and 7 too short, of 11,051 recordings) are excluded before pairing, because digital silence correlates with itself. Added noise, a cough, or a name spliced into a template can only lower the similarity, so the census is a floor on replay, never a ceiling.

\textbf{Speaker embeddings.} Two questions remain: whether the fresh renderings of one script share a voice, and whether one voice serves several scripts. To ask them, we give every window a 256-dimensional speaker vector from a public encoder trained with the generalized end-to-end loss~\cite{wan2018ge2e}. The ``same voice'' threshold is not chosen by hand. Our own ten personas each read different text on hundreds of calls, so pairs of their renders are known same-voice, fresh-render audio; identical-waveform pairs and random pairs bound the scale from above and below. The thresholds are re-derived for each population the check runs on.

On the listening pool, the cosine that passes 95\% of persona pairs is 0.826 (passing 2.1\% of random pairs), with a stricter cut at 0.93 (0.3\%). Over every recorded call the same calibration gives 0.776 (5\% of random pairs) and a strict cut of 0.925, which passes 0.07\% of random pairs and three quarters of persona pairs; §4.7 counts voices at that strict cut. The encoder was trained on wideband speech and these windows are narrowband and noise-suppressed, so the thresholds hold within this corpus only.

\textbf{Call shapes.} An energy-based voice-activity pass over both tracks of every call gives call length, seconds of caller and persona speech, and who spoke first. It also counts the separate runs of caller speech, where segments closer than two seconds form one run. Four shapes follow. \emph{Silent}: under 0.5 s of caller speech. \emph{One shot}: at most two runs and under 20 s of caller speech, with the call ending within 15 s of the last run. \emph{Short exchange}: at most three runs and under 60 s. \emph{Conversation}: everything else. Every table built from these instruments ships with the analysis code, with caller numbers hashed.

\section{Results}

The denominator of this section is every recorded inbound call, not the two-turn corpus. Over 66 days the honeypot recorded 10,987 calls on its numbers (test-line calls removed). The 2,711 calls on eleven outage days follow the rule of §3.1 and sit outside the decomposition below.

On the other 55 days the persona greeted the caller on 7,233 of 8,276 calls (87.4\%), and §4.1 decomposes those 7,233. The detector scored 6,192 calls, all of them inside the two-turn corpus. §4.2 onward report detector-labeled rates over those 6,192 unless noted, because the detector is the only instrument that reaches inside audio that is fresh on every call. Of the 6,192, 318 are calls the persona did not greet (187 on outage days, 131 on normal days); including them leaves the rate unchanged at 29.3\%.

Throughout, ``AI-labeled'' (equivalently, ``flagged'') means detector-labeled as synthetic at the operating threshold, and ``human-labeled'' means not so labeled. Where the replay census separates them, we write \emph{fresh-synthetic} (999 calls) and \emph{fresh-human} (4,180) for the AI- and human-labeled scored calls whose opening is not a replay, and \emph{replayed} (1,013) for the rest. ``AI-labeled'' without qualification always means all 1,816. Excluding the 427 corpus calls that yielded no scorable utterance is not neutral. Assigning all of them to one side moves the detector's headline between 27.4\% and 33.9\%, a bound comparable to the threshold sweep in §4.2.

\begin{figure*}[t]\centering\includegraphics[width=\textwidth]{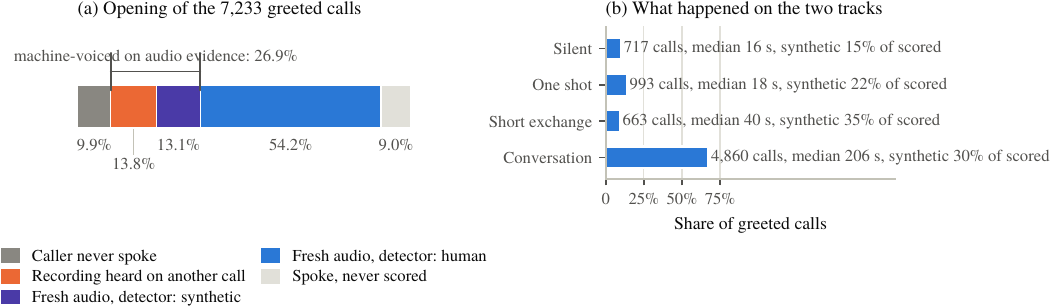}
\caption{\textbf{What the 7,233 greeted calls opened with, and what shape they took.} \textbf{(a)} Every call in one row of Table~\ref{tab:decomp}: silent caller, a recording heard on another call, fresh audio the detector labels synthetic or human, or spoke-but-unscored. The bracket marks the 26.9\% that is machine-voiced on audio evidence alone. \textbf{(b)} Call shapes from the two tracks (Table~\ref{tab:shapes}), with the detector's synthetic share among the scored calls of each shape.}\label{fig:decomposition}\end{figure*}
\subsection{How much of the traffic is a machine}

Three instruments with different blind spots answer this together (§3.5). An audio fingerprint finds the same recording played on different calls; it needs no detector and cannot say whether the recording was made by a person or a synthesizer. The commercial detector reaches inside audio that is fresh on every call, and §4.2 measures how far to trust it. A voice-activity pass over both tracks counts the callers who never spoke at all. We apply them in that order, so every call falls in exactly one row of Table~\ref{tab:decomp} (Figure~\ref{fig:decomposition}).

\begin{table}[t]\centering\small\setlength{\tabcolsep}{1.5pt}
\caption{Decomposition of the 7,233 calls the persona greeted on normal days. A call is classed silent first, then replayed if its opening is the same recording as another call's (alignment similarity at least 0.70, §3.5), then by the detector's label on fresh audio. Intervals are cluster-bootstrap percentiles on originating number (4,364 numbers, 4,000 resamples).}\label{tab:decomp}
\begin{tabular}{@{}>{\raggedright\arraybackslash}p{\dimexpr\columnwidth-4.4em*3\relax}rrr@{}}\toprule
Opening of the call & Calls & Share & 95\% CI \\ \midrule
Caller never spoke after our greeting (under 0.5 s of caller speech; median call 16 s) & 717 & 9.9\% & 9.0--10.9\% \\
Recording also played on another call (audio fingerprint, §3.5) & 995 & 13.8\% & 11.3--16.7\% \\
Fresh audio the detector labels synthetic (maximum per-utterance score at or above 0.85) & 949 & 13.1\% & 12.0--14.3\% \\
Fresh audio the detector labels human (score below 0.85) & 3,921 & 54.2\% & 51.9--56.4\% \\
Caller spoke but was never scored (outside the two-turn corpus, or no scorable utterance) & 651 & 9.0\% & 8.1--9.9\% \\
\bottomrule\end{tabular}\end{table}

Two rows are \emph{machine-voiced}, our term for an opening that is a recording or a synthesized voice, on the audio evidence alone. On that evidence \textbf{26.9\%} (95\% CI 24.3--29.8\%) of greeted calls open with a recording heard on another call or with fresh speech the detector labels synthetic. A third row is almost certainly machine-placed. A caller who stays on the line for a median of 16 seconds after ``Hello?'' without saying anything is what a predictive dialer produces when no agent is free.

A predictive dialer places more calls than it has agents and drops the surplus. It is also what a bot produces when its trigger never fires. A person who wanted the call says something. We keep that 9.9\% in its own row because there is no voice to classify. The remaining 9.0\% could not be scored and is undetermined. The 54.2\% the detector calls human carries the detector's false-negative rate, which §3.4 bounds from below and which the 23 unflagged clips of §4.2 can only bracket.

The two machine rows are different populations. Where the detector scored them, replayed openings are 80.7\% AI-labeled against 19.3\% of fresh openings, so \textbf{45\% of everything the detector flags is a recording that also opened another call}. The label is not uniform across groups: the largest replay group, 85 calls of one recorded pitch, is 0\% AI-labeled. That is the share of the headline synthetic rate that a replay-detection pass, rather than a synthesis detector, accounts for. Whether those recordings were made by a person or rendered once by a synthesizer is what neither instrument can tell; the detector's label on them says that most sound synthesized.

\textbf{The split depends on the fingerprint threshold; the total does not.} Table~\ref{tab:decomp} counts a call as replayed at the 0.70 criterion of §3.5. At the 0.85 criterion, which the calibration certifies as the same file, the replay row falls to 666 calls (9.2\%). The 329 calls it releases move to the fresh-audio rows: 258 to the synthetic row, 51 to the human row and 20 to the unscored row. The synthetic row then reads 1,207 calls (16.7\%), the machine-voiced total 25.9\% against 26.9\%, and the split between playback and synthesis roughly one to two instead of one to one.

The 0.70 groups are also the less certain ones: 96 of the 239 replay groups have a median pairwise similarity below 0.80. And 53 groups (263 calls) share an opening of four words or fewer, where short generic audio can be merged by the census. We report the 0.70 census because the calibration places freshly rendered speech at 0.2--0.5 and the same file at 0.85--0.98, so a pair at 0.70 is far from fresh speech. Readers who want only the certified criterion should read 9.2\% for playback and 16.7\% for synthesis.

\textbf{Call shapes.} Table~\ref{tab:shapes} classes the same 7,233 calls by what happened on the two tracks. Two thirds became conversations. One in seven ended after at most two short runs of caller speech, a median of 18 seconds into the call. That shape fits a recorded pitch, a voicemail drop, or a bot that hung up when the persona's answer was not the one it wanted. Replays and detector flags concentrate in the short exchanges rather than in those one-shot calls. Of short exchanges, 19.5\% open with a recording heard on another call against 10.7\% of one-shot calls, and the one-shot calls are the least often flagged of the calls that spoke at all. Both comparisons are within scored calls; the scored share differs elevenfold across shapes (Table~\ref{tab:shapes}).

\begin{table}[t]\centering\small\setlength{\tabcolsep}{1.5pt}
\caption{What happened on the two tracks, same 7,233 calls. ``AI-labeled'' is among the calls the detector scored; the scored share is 8.5\%, 68.7\%, 84.0\% and 94.1\% down the rows.}\label{tab:shapes}
\begin{tabular}{@{}>{\raggedright\arraybackslash}p{\dimexpr\columnwidth-4.4em*4\relax}rrrr@{}}\toprule
Shape & Calls & Share & Median length & AI-labeled \\ \midrule
Silent: under 0.5 s of caller speech & 717 & 9.9\% & 16 s & 14.8\% \\
One shot: at most two runs of caller speech, under 20 s, then hang-up & 993 & 13.7\% & 18 s & 22.0\% \\
Short exchange: at most three runs, under 60 s & 663 & 9.2\% & 40 s & 35.2\% \\
Conversation & 4,860 & 67.2\% & 206 s & 29.9\% \\
\bottomrule\end{tabular}\end{table}

\textbf{The calls the corpus filter dropped.} The two-turn filter that defines the corpus keeps 6,359 of the 8,276 normal-day calls; 72 more are recordings with no matching session, and the 1,845 the filter drops are not prerecorded blasts. Of those 1,845, 63\% are silent and 22\% are one shot; their median length is 18 seconds, and only 2.1\% open with a recording heard on another call. They are calls on which nobody, or almost nobody, spoke to us, and where their audio can be examined it looks like an abandoned dialer connection rather than a robocall. None of them carries a detector score. That is why the decomposition is reported over greeted calls and not over the corpus: the corpus filter removes the silent machine, not the talking one.

\begin{figure}[!t]\centering\includegraphics[width=\columnwidth]{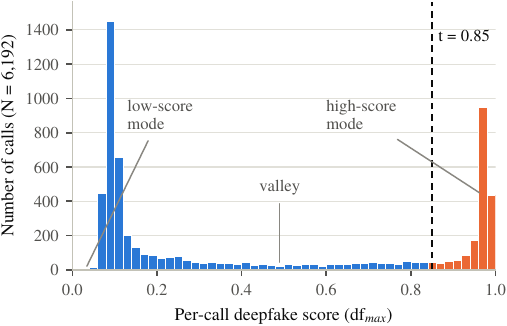}
\caption{\textbf{Per-call detector scores are sharply bimodal} ($n=6{,}192$; one maximum per-utterance score per call). Most calls sit near 0 or near 1 with a sparse middle. Our 0.85 threshold nonetheless sits on the rising shoulder of the upper mode, not at the trough near 0.45--0.50; a threshold at the trough would report 39.6\% rather than 29.3\%, so 0.85 is the conservative end. Bimodality shows the detector separates \emph{something} sharply, not that the axis is synthesis.}\label{fig:bimodal}\end{figure}
\begin{figure}[!t]\centering\includegraphics[width=\columnwidth]{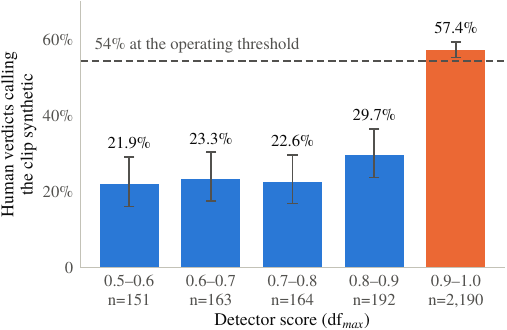}
\caption{\textbf{The detector ranks well and calibrates poorly.} Share of decisive judgments calling a clip synthetic, by detector score band (2,860 decisive of 3,188 judgments; bars weighted by judgment with Wilson 95\% intervals, per-clip averages run 14.6\% to 55.5\%). Confirmation rises from 21.9\% in the 0.5--0.6 band to 57.4\% in the 0.9--1.0 band; the 0.8--0.9 band straddles the 0.85 operating threshold. The below- to above-threshold contrast is positive within every listener who has at least fifteen decisive judgments on each side (eight of eleven). At the operating threshold only 54.4\% (95\% confidence interval 51.5--57.3\%) of the 882 flagged clips are confirmed.}\label{fig:validation}\end{figure}
\begin{figure}[!t]\centering\includegraphics[width=\columnwidth]{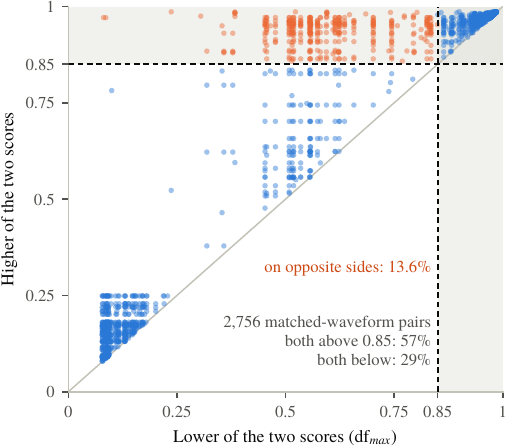}
\caption{\textbf{The same waveform, scored twice.} Each point is a pair of calls whose opening waveforms match, plotted by the lower and the higher of the two detector scores; matched means the alignment similarity of §3.5 is at least 0.85. Shaded bands mark the 0.85 threshold; orange points are pairs the threshold splits, 13.6\% of the 2,756 pairs.}\label{fig:pairs}\end{figure}
\begin{figure}[!t]\centering\includegraphics[width=\columnwidth]{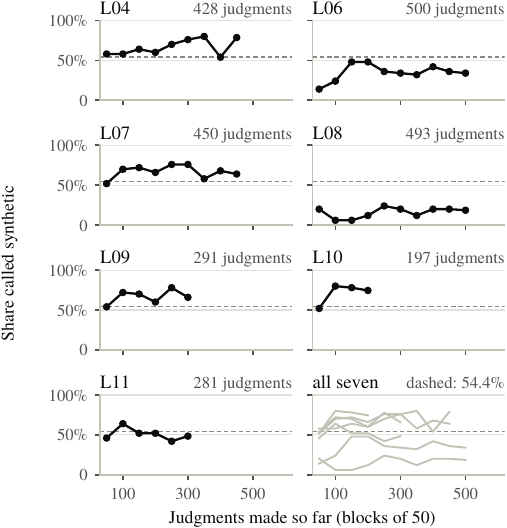}
\caption{\textbf{A listener's criterion wanders.} Share of decisive judgments calling a clip synthetic, in blocks of fifty judgments in the order each listener made them, one panel per listener for the seven with at least a hundred decisive judgments; the last panel overlays all seven. The dashed line marks, for reference, the 54.4\% per-clip confirmation of §4.2 for clips at or above the operating threshold; the panels include the served clips below it. Later blocks meet a harder mix of re-served contested clips, so the level, not the slope, is the point.}\label{fig:drift}\end{figure}
\subsection{Prevalence and threshold stability}

At the operating threshold, 1,816 of 6,192 scorable calls, \textbf{29.3\%}, are labeled synthetic. Treating calls as independent gives a Wilson interval of 28.2--30.5\%, but calls cluster within 3,814 originating numbers. The mean cluster size is only 1.62, yet the distribution is heavy-tailed: the Kish effective cluster size $\sum m^2 / \sum m$ is 6.45, and a bootstrap resampling numbers rather than calls gives a design effect of 4.9. We therefore report the cluster-bootstrap percentile interval, \textbf{95\% confidence interval (CI) 27.0--31.9\%}.

The score distribution is strongly bimodal: 1,909 of 6,192 calls (30.8\%) score below 0.10 and 1,703 (27.5\%) at or above 0.90, with a sparse middle (Figure~\ref{fig:bimodal}, drawn at 0.02-wide bins). Bimodality shows the detector separates some axis of this data sharply. It does not establish that the axis is synthesis, and we do not argue otherwise; codec, bandwidth and trunk class would each produce a confident bimodal split.

Varying the two analyst choices, threshold and aggregation rule, moves the estimate. Taking the maximum per-utterance score gives 39.6\% at a 0.50 cutoff, 29.3\% at 0.85 and 27.5\% at 0.90, while taking the mean gives 32.8\% and 21.8\% at 0.50 and 0.85. The estimate therefore ranges from \textbf{21.8\% to 39.6\%}, a factor of 1.8. We report that full range rather than the narrower max-aggregation window, because the aggregation rule is as much an analyst choice as the threshold.

Human listeners confirm about half of what the detector flags. Eleven listeners supplied 3,188 blinded judgments on the same ten-second windows the detector scored; 2,860 are decisive and the remaining 328 are ``unsure.'' Every judgment joins to a scored call through the released clip-to-call crosswalk. Over the 882 labeled clips at or above the operating threshold, cut from 878 calls, the mean per-clip synthetic fraction is \textbf{54.4\% (95\% CI 51.5--57.3\%)}, bootstrapping clips. Resampling the 644 originating numbers the clips come from instead gives 48.4--60.0\%, since replayed scripts put up to 37 labeled clips on one number. Two pre-specified sensitivity analyses move it little. Excluding the one listener who agrees with the majority of their peers less than half the time gives 58.1\% (55.0--61.0\%). Restricting to judgments where at least five of the ten seconds were actually heard gives 52.6\% (49.1--56.0\%).

Taking 54.4\% as the confirmed fraction of the 1,816 flagged calls puts \textbf{human-confirmed synthetic openings at 15.9\% of scored calls}, under a transfer assumption the pool does not license. The labeled clips over-represent high scores (§3.4) and confirmation rises with score (below), so 54.4\% slightly overstates precision on the census. Reweighting the labeled clips to the census's score mix in two bands split at 0.95 gives 53.2\%, or 15.6\% of scored calls; four bands give 52.8\% and 15.5\%.

With no known-human controls the listeners' own false-positive rate is unmeasured, so even that describes the protocol's estimate, not the true rate. And it is a partial estimate rather than a prevalence because the complementary quantity, the share of \emph{unflagged} calls a listener would call synthetic, currently rests on 23 labeled clips, five of which listeners call synthetic. The Wilson interval on that rate, 0.10--0.42, moves prevalence between 23\% and 46\%: a bracket that contains the detector-labeled 29.3\% without confirming it, and too wide to report as an estimate.

The detector's score is nonetheless informative rather than arbitrary. Human confirmation rises across score bands, from 21.9\% in the 0.5--0.6 band to 57.4\% in the 0.9--1.0 band, with a flat stretch below 0.8 (Figure~\ref{fig:validation}; bands of 0.1 from 0.5, weighted by judgment). Averaged per clip as in §3.4, the same series runs 14.6\%, 16.8\%, 17.8\%, 27.2\% and 55.5\% over 61, 75, 78, 90 and 841 clips. The four lower bands differ by a point or two on 61 to 90 clips each, so the rise is carried by the top band. The point that rules out a composition artifact is that the contrast holds within listeners. For each of the eight listeners with at least fifteen decisive judgments on each side, confirmation is higher above the operating threshold than below it within the served pool (0.50--0.85), by 8 to 44 percentage points.

Listeners disagree with each other about which clips are synthetic (mean pairwise Cohen's $\kappa$ = +0.267) yet agree about which direction the detector's score points. The detector therefore ranks well and calibrates poorly. A score of 0.97 means a clip is far more likely synthetic than a score of 0.6, but it does not mean the clip is synthetic.

\textbf{Listeners drift.} The judgment export carries a timestamp per judgment, so each listener's verdicts can be read in the order they were made (Figure~\ref{fig:drift}). Over blocks of fifty decisive judgments, a listener's share of ``synthetic'' verdicts wanders by ten to twenty points within a few hundred clips. One listener moves from 58\% to 79\%, another from 14\% to 34\%, a third from 46\% to 64\% and back to 48\%.

Six of the seven listeners with at least a hundred decisive judgments trend upward over their whole sequence. But the priority tiers re-serve contested clips later, so later judgments meet a harder mix. Restricted to each clip's first judgment by anyone, the direction is mixed: one listener falls from 58\% to 52\%, two rise, one is flat. What the data support is not that listeners learn, but that a listener's criterion is unstable. The same ear calls a different share of the same kind of audio synthetic depending on when it is asked.

\textbf{The same waveform, two scores.} The replay census supplies a check no listener can (Figure~\ref{fig:pairs}). Of scored calls, 2,756 pairs open with a matched waveform: alignment similarity at least 0.85 over the fifteen-second census window, which contains the ten seconds the detector scored (§3.5). Any difference between their two scores therefore belongs to the instrument and the line, not to the voice. The median absolute difference is 0.008, but the 90th percentile is 0.28, and \textbf{13.6\% of identical pairs sit on opposite sides of the 0.85 threshold}.

Among the 69 replay groups with three or more scored members, 11.6\% contain calls both above and below 0.50. Line, codec and noise alone carry the detector across its own decision boundary on roughly one matched pair in seven. That is a floor on its per-call error that the listening study, which works one clip at a time, cannot see. It is also the reason a call-level label should be read as a draw from a distribution rather than a verdict.

Recorded compliance banners do not carry the rate. Of scored calls, 590 (9.5\%) contain a ``this call is being recorded'' utterance in the scored window. Only 101 of the 1,816 flagged calls have that banner as their top-scoring utterance, and 78 calls cross 0.50 on banner utterances alone. Banner utterances score lower on average (0.32) than other speech (0.39).

\subsection{Automation tracks volume, not harm}

Partitioning by the label our scam-versus-spam classifier assigns inverts the intuitive ordering, in which fraud would be the most automated class (Table~\ref{tab:classes}). Spam, meaning unsolicited but non-fraudulent commercial calling dominated by scripted lead generation, is the most automated class, while calls the classifier labels fraud are markedly less so. Under the narrower credential-request criterion for fraud (698 calls), the share is 15.9\%.

\begin{table}[t]\centering\small\setlength{\tabcolsep}{1.5pt}
\caption{Synthetic-voice share by call class (n = 5,962 classified calls), with cluster-bootstrap intervals on originating number.}\label{tab:classes}
\begin{tabular}{lrrr}\toprule
Class & Synthetic / total & Share & 95\% CI \\ \midrule
Spam & 1,291 / 3,818 & 33.8\% & 30.6--37.2\% \\
Scam & 192 / 910 & 21.1\% & 17.5--25.1\% \\
Unsure & 179 / 876 & 20.4\% & 17.2--23.9\% \\
Legit & 74 / 358 & 20.7\% & 11.5--31.5\% \\
\bottomrule\end{tabular}\end{table}

The rows sum to 5,962 rather than 6,192 because 230 scored calls carry no class label; those are themselves 34.8\% synthetic, so their omission does not manufacture the inversion. We read the inversion as consistent with synthesis adoption following volume economics rather than the severity of the offense. A lead-generation pitch runs a fixed script and benefits when the marginal cost of one more call approaches zero, while extracting credentials from a resistant target requires improvisation. Intervals are cluster-bootstrap percentiles on originating number, so the inversion survives clustering. The legit row's width shows it is the least reliable, mixing disclosed business voicebots with classifier error. The vertical the numbers were seeded into shows through in the scripts. Callers who reference a form for burial or final-expense coverage account for 71 scored calls from 51 numbers, 76\% of them AI-labeled and read by several voices (speaker embeddings, §3.5).

There are three competing readings we cannot exclude. The comparison is unadjusted for call length, which §4.6 shows is confounded with the AI label. It is also unadjusted for exposure age, which §4.4 shows moves prevalence further than length does, so a class mix that differs across our eleven numbers would produce this pattern. And if the detector partly measures non-liveness, prerecorded blasts would concentrate in spam for reasons unrelated to TTS.

Call endings separate AI-labeled from human-labeled calls as sharply as call class does. Calls that end with the caller transferring us to another agent on their side are 69.9\% AI-labeled (256 of 366), against 26.8\% among the 5,826 other calls. Another 45 of those other calls carry no ending label and are counted here as non-transfer. That is an odds ratio (OR) of 6.36 (cluster bootstrap on originating number, 95\% CI 4.90--8.34), rising to 7.29 once duration and turn count are adjusted for. The transfer pattern is not simply an IVR front end handing to a person. Only 1 of the 366 transfer-ending calls opens with a robocall or IVR pattern, and 338 open conversationally as a cold pitch, a reference to prior action, or an authority impersonation.

The ending label is produced by an LLM reading transcript text written by the persona's own ASR, not by the detector API, so the two are separate instruments. They are not fully independent, however. The labeler is also given the caller number. Of the 854 numbers placing two or more calls, 78.1\% are pure with respect to the voice label: every call from the number carries the same label. The ending label could therefore in principle exploit the number. Re-running the labeler with the number withheld would settle it (§5.5).

\begin{figure}[!t]\centering\includegraphics[width=\columnwidth]{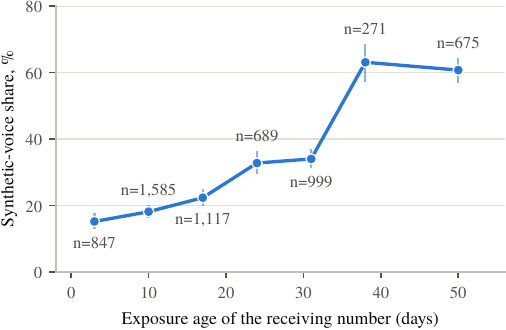}
\caption{\textbf{The apparent trend is our instrument aging, not the ecosystem changing.} Synthetic-voice share of 6,183 of the 6,192 scored calls by the \emph{exposure age} of the number that received each one, the days since that number's first inbound call (nine calls on a twelfth, low-volume line are omitted). Numbers were seeded at different times, so exposure age is not calendar time. Share rises from 15.2\% in the first week to 60.7\% after six; bars are Wilson 95\% intervals and labels give calls per week of age.}\label{fig:exposure}\end{figure}
\subsection{Prevalence tracks a number's exposure age, not calendar time}

The weekly synthetic share rises steeply across our collection window. It is not an ecosystem trend. Our numbers were seeded at different times over about a month, so a number's exposure age and the calendar date run on different clocks, and the rise follows the aging of our own bait numbers (Figure~\ref{fig:exposure}).

A logistic model containing both terms gives exposure age β = +0.0505 per day against calendar date β = −0.0004 per day. Fitted alone, the age model reaches an Akaike information criterion (AIC, lower is better) of 6,888 against 7,372 for calendar date, and the model carrying both sits at 6,889.7, indistinguishable from age alone. We report these as descriptive fit statistics only, and attach no $p$-value to either coefficient, for a reason the design makes plain.

Exposure age varies at the level of the receiving number, of which there is exactly one before 1 July 2026. All separation of age from date comes from the twenty-day window in which eleven numbers coexist. The ten fresh numbers were seeded on a single day into the same funnels, so they behave as one cluster. The effective number of independent units is closer to two than to 6,192, and no $p$-value computed as though calls were independent draws is meaningful here.

The natural experiment carries the section. The window runs from the first call to a fresh number on 1 July to 20 July, the last full day of collection, attributing each call to the number that received it. Over it the ten fresh numbers received 18.6\% synthetic-voiced calls (n = 1,759). The number seeded in May stood at 59.4\% (n = 1,088) over the same days. Matched on exposure age rather than date, the contrast disappears. Over its own first twenty days of exposure the May number ran at 18.4\% (n = 1,650, median call date 8 June), against 18.6\% for the fresh numbers over theirs (median 13 July). That is a difference of +0.2 percentage points across 34 days of \emph{seeding} separation. We report the contrast as an effect size and attach no test: with two effective clusters the interval is far wider than the ±3 points an independence assumption would give.

The apparent rise lives in traffic presenting a caller ID in the 484 area code, the bait numbers' own, so that the call looks local (neighbor spoofing). The remaining 4,115 calls, 70\% of the calls in the seven full collection weeks, move only from 16.2\% to 19.7\% (Spearman ρ = 0.46 on the weekly series). Seven weekly observations are too few to detect a trend of the size at issue.

The regression and the two-group contrast, derived independently, agree. The fitted slope of 0.0505 per day implies that the 18.6\%-to-59.4\% gap corresponds to about 37 days of age separation, against an actual seeding separation of 34 days.

The mechanism we infer is lead resale: a seeded number circulates down the chain and is progressively acquired by larger, more automated operations. Three caveats bind it. The labels are detector-assigned. The portion of the age curve beyond roughly twenty days rests on a single number. And lead resale is inferred from the correlated rise in spoofed caller ID, not observed.

The consequence for the headline is direct: 29.3\% averages over our own seeding schedule. The same number, once it had circulated for over a month, was running at 59\%, while ten numbers seeded that July ran at 19\% over the same days.

Volume is a separate matter from share. Over the same twenty days the May-seeded number drew 54 scored calls a day against about 9 for each new number, and it had drawn 83 a day in its own first twenty days. This section is about the share of those calls that are machine-voiced, not about how many arrive, which the companion analysis~\cite{traister2026anatomy} reports.

\begin{figure}[!t]\centering\includegraphics[width=\columnwidth]{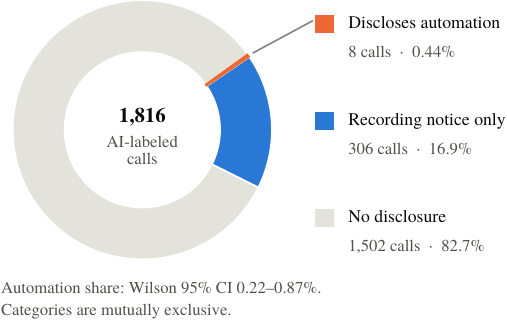}
\caption{\textbf{Callers will say a call is recorded; they will not say it is automated.} Of the 1,816 AI-labeled calls, 8 (0.44\%) disclose that the call or system is automated, 306 (16.9\%) announce only that it is recorded or monitored, and 1,502 (82.7\%) do neither. Table~\ref{tab:disclosures} gives the 8 verbatim. The banner ``AI is taking notes'' is excluded: it announces a notetaker, not the speaker.}\label{fig:disclosure}\end{figure}
\subsection{Automated callers almost never announce themselves}

Measuring disclosure requires separating two acts that keyword matching conflates. A genuine self-disclosure states that the present call or system is automated. TCPA autodialer-consent boilerplate, a request for permission to autodial on \emph{future} contacts embedded in telemarketing consent scripts, is a compliance act, not an admission about who is speaking now. Distinguishing them changes the count: 37 of the 42 apparent disclosures in the human-labeled population are a single autodialer-consent template, read under several insurance brands.

Under the corrected construct, genuine self-disclosure occurs on \textbf{8 of 1,816 AI-labeled calls, 0.44\%} (Wilson 95\% CI 0.22--0.87\%). Self-disclosure is therefore close to absent: 99.6\% of calls whose opening we label synthetic never say so.

One phrase is deliberately excluded. The banner ``AI is taking notes'' opens 14 further AI-labeled calls (a fifteenth scores below threshold, §3.4), and a keyword approach would count it. But it names an AI notetaker transcribing the conversation rather than describing the call or the system placing it. It is a statement about a tool in the room, not about who is speaking, and counting it would nearly triple the rate. The first two banners of Table~\ref{tab:disclosures} are counted. They attribute AI to the call itself (``this call uses AI''; ``this call is being … processed and analyzed by AI'') rather than to a named third-party tool. A reader who treats those as tooling statements too would put the rate at 5 of 1,816 (0.28\%, Wilson 95\% CI 0.12--0.64\%).

\begin{table}[t]\centering\small\setlength{\tabcolsep}{1.5pt}
\caption{Every genuine self-disclosure among the 1,816 detector-flagged calls, verbatim. Bracketed terms replace the caller's stated principal, which we cannot verify and which may be impersonated.}\label{tab:disclosures}
\begin{tabular}{@{}>{\raggedright\arraybackslash}p{\dimexpr\columnwidth-4.4em*1\relax}r@{}}\toprule
Disclosure, verbatim & Calls \\ \midrule
``This call is being recorded, processed and analyzed by AI.'' & 2 \\
``This call uses AI and will be recorded. Let's get started.'' & 1 \\
``I'm Emily, a virtual assistant with [roofing firm].'' & 1 \\
``I'm an AI assistant from [insurance agency].'' & 1 \\
``Welcome to [government program]'s automated customer survey.'' & 1 \\
``Welcome to [security company]'s automated secure payment platform.'' & 1 \\
``Yes. I am. I'm a robot.'' & 1 \\
\bottomrule\end{tabular}\end{table}

Disclosure is a property of the tooling rather than of the operator. Every entry but the last is a scripted banner a platform emits before the conversation begins. Several name a business or government program as the caller's principal, a claim we cannot verify and therefore mask. The one volunteered admission, ``Yes. I am. I'm a robot,'' came only after our persona asked the caller directly whether it was one. Nobody in this corpus discloses unprompted except software configured to.

Disclosure of \emph{recording}, by contrast, is routine. Of the same 1,816 calls, 306 announce that the call is recorded or monitored and never that it is automated; a further 3 do both, in the first two banners of Table~\ref{tab:disclosures} (Figure~\ref{fig:disclosure}). Callers are evidently willing to read a compliance preamble. What they omit is the part about who is speaking.

We report no human-labeled comparison. A caller that states it is automated and scores below our threshold is a detector false negative, not a candid human. A comparison group would therefore measure the detector's errors while appearing to measure human behavior; such calls are counted instead as a floor on the false-negative rate (§3.4). Establishing the 8 required a sweep over all 189,555 transcribed caller turns in the corpus message log with hand adjudication of the 82 candidates it surfaced. The 8 disclosures come from 7 distinct caller numbers, so they are not one campaign, but a count this small supports no comparison and we attach none.

\subsection{Behavioral signatures, and four nulls}

Calls the detector labels synthetic open with a robocall or IVR pattern far more often: 13.33\% (242 of 1,816) against 2.93\% (128 of 4,376). The unadjusted odds ratio of 5.10 attenuates only to 4.71 once duration and turn count are adjusted for, and the effect survives clustering on caller number ($p = 9.9\times10^{-4}$, marginal under the correction noted in Appendix A).

AI-labeled callers run shorter, thinner conversations, median 104.0 s against 174.5 s and 8 caller turns against 13, but most of that gap is compositional. Reweighting human calls to the AI mix of originating-number classes absorbs 63\% of the median duration difference. The residual is carried by toll-free traffic, and is absent in the other two originating-number classes, the local 484 spoofed block of §4.4 and other geographic numbers, which together account for 81\% of calls. On most of this corpus there is no duration difference to explain, and neither the stratified duration nor the turn-count comparison survives correction for the roughly fifty comparisons this paper reports (Appendix A).

Opening lines recur near-verbatim across separate calls from the same originating number when that number is AI-associated. Median cross-call opening Jaccard is 0.735 across 65 pure-AI numbers against 0.188 across 602 pure-human ones, while within a single call there is no difference ($p = 0.117$). What distinguishes automated calling here is reuse of a fixed opener from call to call, not a loop inside one conversation. Three cautions apply. The detector labels calls, not numbers, and 187 numbers carry calls of both labels. Pure numbers are lower-volume, so this runs on a selected subsample. And text similarity alone cannot distinguish a script re-rendered from a file replayed, which is what the audio fingerprint of §4.1 adds.

\textbf{Answering beside the point.} A bot that cannot hear the persona answers a question with the next line of its script. From the two-sided transcript we take every persona turn that ends in a question. A reply counts as beside the point when it shares no content word with the question, carries no yes, no or number, and still runs on for eight or more words.

\begin{figure*}[t]\centering\includegraphics[width=\textwidth]{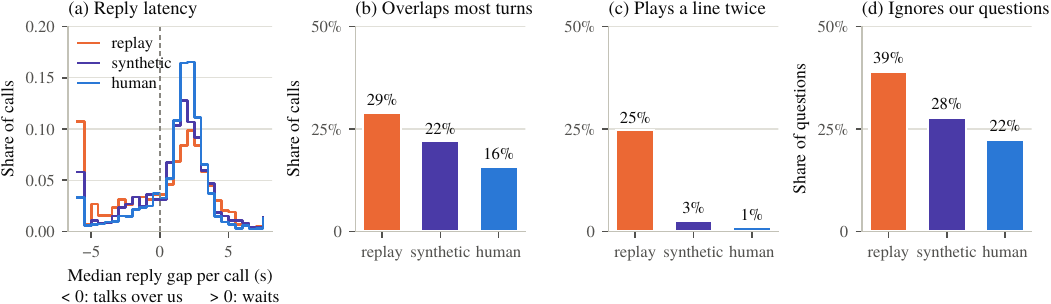}
\caption{\textbf{Machine behavior inside a single call.} \textbf{(a)} Per-call median gap between the end of our persona's turn and the caller's next run of speech, by class; negative gaps are callers talking over us. \textbf{(b)} Share of calls whose caller overlaps the persona on more than half of its turns. \textbf{(c)} Share of calls in which the same recording is played twice within the call (alignment similarity of §3.5 at least 0.70 between two runs of caller speech). \textbf{(d)} Mean share of the persona's questions left unanswered (§4.6). Every signal needs one call and no corpus.}\label{fig:conversation}\end{figure*}

By that rule AI-labeled callers leave 32.8\% of our questions unanswered against 22.7\% for human-labeled callers, and 29.7\% of AI-labeled calls ignore more than half of the questions asked, against 14.9\%. Split three ways it is 39.0\% for replayed openings, 28\% for fresh-synthetic and 22\% for fresh-human (Figure~\ref{fig:conversation}d), widest where the voice is a recording, as it should be. Near-verbatim repetition of a line within one call is the signature of a soundboard, an operator playing recorded lines from a keypad. It is rare everywhere (1.8\% of long caller turns) but eight times more common in replayed calls (5.0\%) than in fresh-audio calls (0.6\%). The rule is crude, and a person who talks over a question is counted the same as a bot; what we report is the direction and the ordering across the three groups.

Four expected findings did not survive. \textbf{Conversion:} apparent scam successes are rarer among AI-labeled calls in raw terms (19 of 1,816 against 109 of 4,376). The difference does not survive control for length (Mantel--Haenszel OR 0.64, $p = 0.068$), and with 19 events we have no power to separate the populations. \textbf{Credential asks:} unadjusted, AI-labeled callers reach a credential request, the criterion of §4.3, far less often (6.11\% against 13.41\%). Adjusting for duration and turn count attenuates this to OR 0.68, and within the scam stratum it vanishes (adjusted OR 0.95, $p = 0.82$). Automated callers do not avoid the ask so much as fail to reach it.

\textbf{A funnel:} it is tempting to join automated openings to transfer endings as an IVR front end handing to a live closer. But exactly 1 of the 370 robocall-or-IVR openings ends in a transfer, against 6.27\% (365 of the other 5,822 scored calls); robocall openings and transfer endings are all but disjoint sets of calls.

\textbf{Two species of automation:} if the 29.3\% concealed both prerecorded blasts and conversational agents, the two should separate behaviorally. Clustering the AI-labeled calls at k = 2 on five conversational features gives a silhouette of 0.326 against a 0.262 noise floor, while the human population clusters slightly better at 0.350. What was tested, and all that is established, is that k = 2 k-means on five features does not recover a separation. The audio does. §4.1 splits the flagged calls into recordings heard on another call (45\%) and audio fresh on every call, a division the conversational features could not see. The robocall-opening enrichment above finds the same automated subgroup in the transcript.

\begin{figure}[!t]\centering\includegraphics[width=\columnwidth]{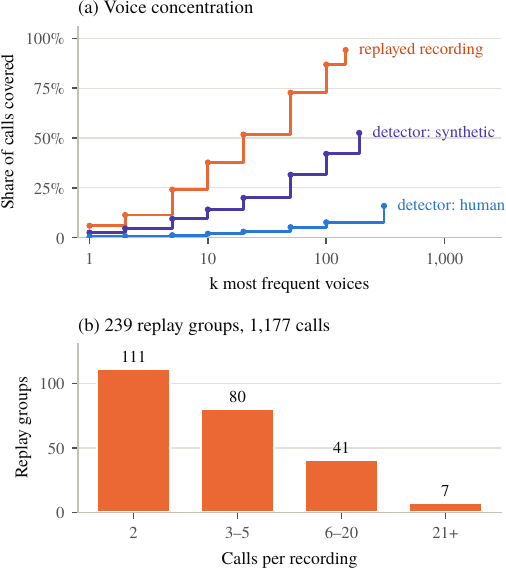}
\caption{\textbf{Voices and recordings.} \textbf{(a)} Share of calls covered by the $k$ most frequent voices, at the strict speaker-embedding threshold of §3.5 (cosine 0.925, passing 0.07\% of random pairs), for fresh audio the detector labels synthetic, replayed recordings, and fresh audio it labels human: synthetic voices recur, human ones almost never do. \textbf{(b)} Size distribution of the 239 replay groups; two thirds are played from two or more caller numbers.}\label{fig:voices}\end{figure}
\subsection{Numbers churn everywhere; recordings and voices persist across campaigns}

We clustered calls by opening script. The first substantive caller turn is normalized by lowercasing, stripping digits and removing personal first names, including our personas' own, which otherwise chain unrelated scripts together through the honeypot's identity spoken back at it. Calls whose token-set Jaccard similarity reached 0.50 are then grouped under single-linkage clustering. Of 6,192 calls, 5,516 yield a usable opener. Single-linkage chains, and it chained here: the largest component held 656 calls at mean internal similarity 0.052, an order of magnitude below the 0.50 threshold that assembled it. We excluded it, leaving the \textbf{420 campaigns} of two or more calls that every figure below is computed over. That exclusion is a post-hoc judgment; a non-chaining method with a uniform cohesion rule would be better. A script that branches on the callee's answer, reading the easy name and confirming the hard one (§4.8), can also split across clusters: we normalize names but not branches, which makes campaigns smaller, never larger.

Against these campaigns, the number is not the durable object. Among the 300 of those 420 campaigns dialed from two or more distinct numbers, the campaign's observed span exceeds the longest span of any single one of its own numbers in 293 cases. That direction is close to guaranteed, because a campaign's span can only grow as numbers are added to it, so the sign carries no information. The magnitude has to be read against what the corpus's number supply would produce on its own, which we test below. The median campaign spans 14.3 days while the median longest-lived number inside a campaign spans 0 days: the typical constituent number places all its observed calls within a single day and is never seen again. The sharpest instance is a campaign that placed 68 calls from 68 distinct numbers over 31 days.

A permutation null puts that churn in context. We reassigned caller numbers to calls at random within each calendar day (10,000 permutations, preserving daily volume and every number's call count). A random 68-call cluster over the same days then has 66.5 ± 1.2 distinct numbers, and all 68 distinct in 23\% of draws. The disposability of numbers is therefore a property of the corpus's number supply, in which 2,960 of 3,814 numbers place a single call, rather than something campaigns add.

What campaign membership does add runs the other way. Among the 24 campaigns with at least 20 calls the distinct-number ratio is bimodal. Eleven dial from a fresh number on nearly every call, five dial from one or two numbers for 18--47 days, and eight mix the two. The null never produces the single-number mode ($p < 10^{-4}$), and 13 of the 24 use significantly fewer numbers than random assignment would. The per-call mode is AI-heavy, 82\% of its 459 calls being AI-labeled against 41\% and 39\% in the other two modes, a difference that does not reach significance at $n = 24$ (Mann--Whitney $p = 0.059$). The durable unit is the script; whether the number is disposable depends on the operation, and for a fifth of the large ones a number blocklist would have worked.

Automation does not appear to accelerate number turnover. Among the 854 numbers placing at least two calls, the 65 whose calls are entirely AI-labeled have a median observed lifetime of 7.0 days, against 5.3 days for the 602 entirely human-labeled. The difference runs opposite to the obvious hypothesis and is not significant (Mann--Whitney $p = 0.42$).

\textbf{Repetition predicts the label.} The more often a script recurs, the more often the detector flags it. Among scored calls with a usable opener, 14.5\% of those whose script is seen once are AI-labeled. The share is 32.4\% in campaigns of two to four calls, 49.6\% in campaigns of five to nineteen, and 60.0\% in campaigns of twenty or more (the chained component excluded). Either a script worth repeating is a script worth automating, or an automated script is what gets repeated; the corpus cannot say which.

\textbf{What the audio adds.} Grouping by opening script finds campaigns; grouping by opening waveform finds the assets they are built from. The replay census (§3.5) places 1,177 recorded calls in 239 groups that share a recording. Of those groups, 162 span two or more caller numbers, so the same file is played from many numbers. And 636 of the 1,177 calls sit in no script cluster at all, because a transcript that differs by a name, a number or a transcription error hides the audio identity beneath it. Thirty-four groups span two or more campaigns. The largest of those is not a pitch but a compliance notice. One recording of ``This call will be recorded for quality purposes'' opens 22 calls from 8 numbers in 6 campaigns, the fingerprint of a shared dialing platform rather than of a shared operator.

Speaker embeddings~\cite{wan2018ge2e} add a grouping by voice. Among the 113 script clusters whose members are not identical recordings, 95 are read by a single voice on every call (75 of the 108 such clusters in the listening pool, where the thresholds were first calibrated). That is the pattern of a text-to-speech voice rendering the script afresh each time. The threshold that defines ``single voice'' passes 95\% of pairs of our own personas' renders and 2.1\% of random pairs on that pool (§3.5). At the pool's stricter cut, passing 0.3\% of random pairs, 91 voices serve two or more campaigns: one reads seven different scripts across three campaigns from two numbers, another serves nine campaigns from seventeen numbers. A voice that recurs across unrelated scripts and numbers is a product, not a person.

\textbf{How many voices.} Counting voices needs a threshold strict enough not to merge strangers. Over every recorded call we use the strict cut of §3.5, cosine 0.925, which passes 0.07\% of random pairs and three quarters of our own personas' same-voice pairs. The counts below therefore split some true voices, and are upper bounds on distinct voices (Figure~\ref{fig:voices}). Among the 4,180 fresh-human calls (all scored calls, not only greeted ones, so more than the 3,921 of Table~\ref{tab:decomp}), 3,515 (84\%) carry a voice heard on no other call. A further 308 voices recur, the largest on 19 calls, and the fifty most frequent voices cover 5\% of the calls: thousands of different people.

Among the 999 fresh-synthetic calls, 474 (47\%) carry a voice heard once. Another 190 voices recur, the largest on 25 calls, and the fifty most frequent cover 32\%; among replayed openings the fifty most frequent voices cover 73\%. The partitions agree where they should, measured by adjusted mutual information~\cite{vinh2009ami}, where 1 means two groupings coincide and 0 means chance agreement. Voice and campaign align for fresh-synthetic calls (0.52 over the 483 in campaigns) more than for fresh-human ones (0.29 over 620). Voice and originating number align for fresh-human calls (0.59 over 665) and not for fresh-synthetic ones (0.30 over 525). Voice and opening script align for both (0.72 over 187 and 0.83 over 19 clustered calls). A person calls from their own number; a synthetic voice is dialed from many.

One observation cuts against treating a campaign as an operator: at least one spans the detector's own AI/human boundary, the identical script appearing on calls labeled synthetic and human. Either the detector mislabeled some, or one operation runs the script through both automated and live instances. Our data do not distinguish them, which is why we call these script families, not organizations.

\begin{figure*}[t]\centering\includegraphics[width=\textwidth]{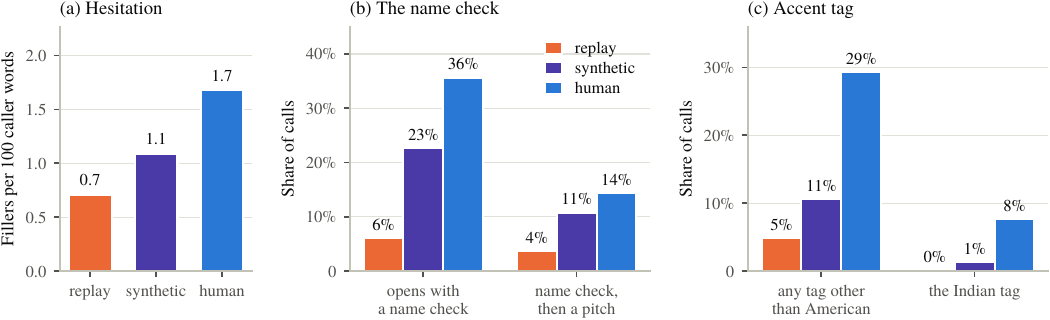}
\caption{\textbf{What the scripts do to sound human.} \textbf{(a)} Fillers per hundred caller words, by caller class. \textbf{(b)} The share of calls that open with a name check, and the share that follow the check with a pitch of fifteen words or more. \textbf{(c)} The detector's accent tag: the share of calls tagged with any accent other than American, and with the Indian tag. The tag comes from the same system as the score, so it is a property of the detector as much as of the callers (Appendix A).}\label{fig:scripts}\end{figure*}
\subsection{What the scripts do, what the channel hides, and what a single call still shows}

Whoever runs an automated calling operation has an incentive to sound like a person, both to the callee and to the detection that carriers and handsets now run. We looked for the tricks that make a call sound human: fillers, name checks that create a small interaction, accents, background noise and degraded audio (Figure~\ref{fig:scripts}). Coughs leave no trace in a transcript and are not counted.

\textbf{Fillers and name checks are in the scripts, but less often than in live speech.} Two thirds of fresh-synthetic calls contain at least one filler (``um'', ``uh'', ``you know''), against 70\% of fresh-human calls and 53\% of replayed ones. The median rate is 1.1 fillers per hundred caller words for fresh-synthetic callers, 1.7 for fresh-human and 0.7 for replayed. Synthesized speech in this traffic is written with some hesitation in it, which is what the speech-synthesis literature recommends for sounding spontaneous~\cite{szekely2019fillers}.

But on every measure we have it hesitates less than the human-labeled callers do, and we have no un-humanized baseline to compare against. The name check is the commoner trick: 22.6\% of fresh-synthetic calls begin by asking who is speaking (``Is this Edna?''). Only 10.7\% follow that check with a pitch of fifteen words or more, the decision-tree pattern in which the easy name is read out and the hard one is confirmed first. Fresh-human callers do both more often (35.6\% and 14.3\%) and replayed recordings almost never (6.0\% and 3.6\%), because a recording cannot wait for an answer.

\textbf{Accents.} The detector API returns an accent tag with each call; Appendix A treats it as a property of the API. Within that caveat, the tag is ``Indian'' on 7.6\% of fresh-human calls and 1.3\% of fresh-synthetic ones, ``American'' on 70.7\% against 89.4\% (95.2\% of replays). Over the two-way split of all scored calls the American shares are 71.7\% and 92.2\% (Appendix A). Whatever the callers intend, the synthesized voices in this traffic carry a generic American tag far more often than the human-labeled ones do; whether that is the voices or the tagger is what Appendix A cannot tell.

\textbf{Background noise and degraded audio cannot be measured here, and that is itself a finding.} The telephony provider applies noise suppression to the inbound leg before it reaches our recorder (Appendix A), and it works. On 8,797 opening windows with speech, the quiet frames sit at digital silence in every class: a median noise floor between −96 and −100 decibels relative to full scale (dBFS). The frequency below which 85\% of the speech energy lies sits near 1 kHz for every class, no class carries energy above 4 kHz, and clipping is absent.

A caller who adds call-center chatter or a crackling microphone to sound like a person loses that noise on the way to us, and so does any detector deployed behind the same suppression. The one channel-level difference that survives is density: a replayed recording talks through 43\% of its opening frames, a fresh-synthetic caller through 31\%, a fresh-human one through 26\%. Recordings do not pause.

\textbf{Two signals that need no second call.} A call center that dials each number once never forms a replay group (Appendix A), so we looked for signs of playback inside a single call, on the two tracks (Figure~\ref{fig:conversation}). The first is turn-taking. For every run of persona speech we measure the gap to the caller's next run. A person waits for the end of our sentence and answers within a fraction of a second~\cite{stivers2009turntaking}; a recording talks over it. Replayed openings answer with a median gap of 1.3 s and begin 40\% of their turns while our persona is still speaking. Fresh-synthetic openings wait 1.6 s and overlap on 36\% of turns; fresh-human openings 1.8 s and 31\%. And 29\% of replayed calls overlap us on more than half of their turns, against 22\% of fresh-synthetic and 16\% of fresh-human calls.

The second is self-replay: the fingerprint measure of §3.5 applied to pairs of caller runs inside one call. A quarter of replayed calls (24.8\%) play the same recording twice within the call, against 2.7\% of fresh-synthetic and 1.1\% of fresh-human calls. One caution applies to both signals. The replayed class is itself defined by the cross-call version of the same fingerprint, and a played file cannot wait for a turn, so part of the separation is the instrument agreeing with itself. The split by the detector's label alone (2.7\% against 1.1\%, 1.6 s against 1.8 s) is the independent part. Together with the answering-beside-the-point rate of §4.6, these are the signals an on-handset screener could compute from one call, without a corpus behind it. None is yet a classifier, and we release the per-call values for anyone who wants to build one.

\textbf{What this adds up to.} On every trick we can measure, the synthesized callers use it less than the human-labeled ones do: they hesitate less and check the name less. The channel removes noise and degradation before any detector sees them. What separates a machine from a person on this evidence is not how it sounds but how it behaves: it talks over us, it repeats itself, and it answers beside the point.

\section{Discussion and Conclusion}

\subsection{What the 2024 ruling did, and what these numbers say about it}

On 8 February 2024 the FCC released a declaratory ruling, adopted unanimously six days earlier, holding that calls made with AI-generated voices are ``artificial'' within the meaning of the Telephone Consumer Protection Act~\cite{fcc2024ruling}. The mechanism is narrower than the headlines it drew. The ruling created no new offense and banned nothing outright. It resolved a definitional question, and in doing so attached the TCPA's existing machinery to synthetic speech: prior express written consent before a telemarketing robocall, identification of the entity responsible, and opt-out on advertising calls. Enforcement runs through Commission forfeitures, carrier-level blocking, a private right of action, and state attorneys general operating under memoranda with the FCC in 48 states. The ruling took effect immediately.

Three of our results bear on it.

\textbf{The regulated conduct is common, but not the conduct the ruling foregrounds.} A quarter of the callers our persona greeted opened with a machine voice on the audio evidence. The split between a recording played on other calls and speech synthesized for the call depends on the fingerprint's own threshold (§4.1). Another tenth never spoke at all (§4.1). The TCPA's own phrase, ``artificial or prerecorded voice,'' covers both halves. The harm is not speculative. But the ruling is framed around voice cloning, meaning impersonating a relative, a celebrity, or a candidate. That is not what an answering honeypot mostly receives: synthetic voice concentrates in lead generation rather than in the fraud stratum (§4.3). A rule argued as an anti-impersonation measure lands, in volume, on telemarketing.

\textbf{The compliance question it actually poses is one this honeypot can only half observe.} The duty the ruling attaches is consent, and we know only half of it. Our fabricated lead checked whatever consent-to-contact box the seeding forms presented (§3.1), so a caller relying on that consent, one that bought the lead without it, and one that never had it are indistinguishable to us. ``Unwanted'' throughout is the recipient's word, not the law's: a call the recipient did not want when it came, whatever box was ticked earlier (§1), and not a call we have shown to be unlawful. What we can measure is a different thing, and the distinction matters.

The TCPA's identification requirement is about the entity responsible for the call, not about the technology producing the voice; there is at present no federal duty to announce that a voice is synthetic. Our 0.44\% (§4.5) is therefore not a compliance rate for a rule that exists. It is a baseline for one that does not, and the baseline is close to zero. The contrast with the 17\% that disclose recording is the useful part: callers already read compliance preambles, so the omission is specific to the speaker, not a general unwillingness to disclose.

\textbf{Enforcement acts on the part of an operation that is cheapest to replace.} Remedies running through per-number reputation and carrier blocking aim at the number, which is discarded within a day, while the script persists for weeks (§5.2).

Read together: the ruling closed a definitional loophole quickly and unanimously, which was worth doing, but its reach is bounded by two things it did not do. It did not require a caller to say it is a machine, and it did not change the unit enforcement acts on.

\subsection{The durable unit of a calling operation is the script, not the number}

Telephony abuse defense is organized around telephone numbers. Blacklists, reputation scores, carrier analytics and consumer call blocking all take the originating number as the thing identified, scored and denied. Our campaign analysis says that is the perishable part for most operations, and, under a permutation null, no more perishable inside a campaign than across the 6,192 scored calls at large (§4.7). The typical number in a campaign places all of its calls inside one day and is never seen again. The campaign itself, the same opening script dialed from a stream of fresh numbers, runs on for a further two weeks.

For that majority, blocking any one number removes almost none of an operation's remaining life; by the time a number earns a bad reputation, the operation has already moved off it. For the minority that hold one number for weeks, five of the 24 largest campaigns and 207 calls, a number blocklist is exactly the right tool. The two modes are told apart by the script, not the number.

Our campaign result sharpens rather than contradicts the field's self-criticism. Prior campaign clustering groups calls by audio similarity and is explicit that it identifies audio, not operators~\cite{prasad2020whoscalling}; ours has the same limit, and one cluster demonstrably spans both sides of the detector's boundary. The practical implication survives: text-level campaign identity is available in real time to anyone already transcribing calls, persists across number rotation, and is not what current blocking acts on. The campaign result holds across the scored calls, not within the synthetic slice, and numbers whose calls are entirely AI-labeled do not turn over faster than human-labeled ones.

\subsection{The honeypot's history shapes the measurement}

Our most transferable methodological finding is a trend we do not report. The weekly synthetic share rises steeply across our window, and it would be straightforward to publish that rise as an ecosystem trend. It is instead a property of our own bait: synthetic share tracks how long a seeded number has circulated rather than calendar date (§4.4).

No amount of resampling removes this confound, because it lives in the history of the honeypot itself, which numbers existed and for how long, not in the data-generating noise. The lesson holds even though our own design identifies the age effect weakly, off what amounts to two independent units (§4.4). Honeypot seeding history is a first-class confounder for any longitudinal claim about telephony abuse, and prior work reporting trends should re-examine its claims for the same structure, particularly where line pools grew or aged unevenly.

\subsection{Two things a listener learns, and a trend}

\textbf{People cannot hear it, and neither can the same person twice.} Listeners confirm about half of what the detector flags and agree with each other only weakly (§4.2), in line with laboratory findings that speech deepfakes are hard to hear~\cite{mai2023humans,muller2022perception}. The listening data add a subtler point: a single listener's criterion moves by ten to twenty points over a few hundred clips, in no consistent direction (Figure~\ref{fig:drift}). A protocol that asks people to label voices should expect that, and should measure a listener against themselves as well as against the instrument.

\textbf{The traffic will move toward machines, because labor is the expensive part.} Of the calls our persona greeted, roughly a seventh each already open with a recording and with synthesized speech. The synthesized voices are concentrated in about two hundred recurring voices serving many scripts and numbers (§4.7). Every incentive points the same way: a recording costs nothing per call, a synthetic voice costs cents, a person costs dollars. And the two-turn filter that older honeypots apply hides exactly the machine-placed calls that never become a conversation (§4.1). We expect the machine-placed share to rise, the replayed share to fall as synthesis becomes cheaper than storing recordings, and the tricks of §4.8 to multiply as detection is deployed at the handset. Measuring that requires the denominator this paper uses, every answered call, and instruments that need no second call.

\subsection{Future work: what would settle the open estimates}

\textbf{Score the calls we could not.} Of greeted calls, 651 spoke but were never scored (Table~\ref{tab:decomp}), and 1,138 of the 1,326 outage-day calls with caller speech remain undetermined (§3.1). Scoring them with the same detector, and listening to a sample of the one-shot calls, settles the undetermined tenth of Table~\ref{tab:decomp} and turns the inference that silent callers are machines into a measurement.

\textbf{Finish the listening study.} Labeling the unflagged stratum and adding known-human gold controls turns the 15.9\% confirmed share of §4.2 into an estimate. A third label, \emph{recording}, with definitions and practice clips before the measured session, would let human ears adjudicate the replay row and would remove the ambiguity the interface left listeners in (§3.4).

\textbf{A second detector.} Running another vendor's detector on the same clips establishes whether 29.3\% is a property of the audio or of one decision surface, and whether the 13.6\% flip rate on identical waveforms is shared.

\textbf{Replay without a second call.} The census needs a recording played to us twice. A word-aligned comparison of shared words would catch templates with a name spliced in. The single-call signals of §4.8 (turn-taking, self-replay, answering beside the point) are candidates for a classifier that needs no corpus, and we release the per-call values for that purpose.

\textbf{The raw line.} The provider's noise suppression removes whatever background a caller adds before it reaches us (§4.8). Recording the inbound leg before suppression would show whether call-center noise and degraded audio are used to sound human, and what suppression does to synthesis artifacts.

\textbf{Later in the call.} Scoring several windows per call would show how often a synthetic opening hands to a person, and how often a person hands to a machine (Appendix A). Re-running the ending labeler without the caller number would show whether that label leaks the voice label (§4.3). And inference clustered on campaign rather than on number would put the right unit under every comparison (Appendix A).

\subsection{Conclusion}

Of the 7,233 greeted inbound calls we analyze, 13.8\% open with a recording we also heard on another call and 13.1\% with fresh audio a commercial detector labels synthetic. A further 9.9\% open with a caller who never spoke, 54.2\% with fresh audio the detector labels human, and 9.0\% could not be scored. Machine-voiced openings are therefore at least 26.9\% (95\% CI 24.3--29.8\%), and, if silent connections are machine-placed, 36.8\%, of what reaches an answering honeypot seeded into lead-generation funnels. The replay half of that figure rests on an audio fingerprint that needs no training data and would survive the retirement of every detector we used. Its split from the synthesis half, though, moves with the fingerprint's own threshold (13.8\% at 0.70, 9.2\% at 0.85).

The synthesis half is detector-conditional, threshold-sensitive across 21.8--39.6\% of scored openings, conditional on our seeding schedule, and human-validated on one side only. Listeners confirm 54.4\% of what the detector flags, the unflagged remainder was not labeled, and the same waveform crosses the detector's threshold on roughly one pair in seven. Ours is also the first such measurement whose thresholds, score distributions, aggregation rules and engagement model are disclosed well enough to be argued with. The one competing public figure, approximately 25\%~\cite{hiya2025}, sits inside the range our choices span. That report discloses too little for us to know what its pipeline shares with ours, and it does not say how much of its 25\% is a recording.

The shape of the population is more informative than the number. Automation concentrates in high-volume lead generation rather than in fraud, and in calls that end with a transfer to another agent on the caller's side. It also concentrates in callers who answer our questions with the next line of a script. Differences we expected, such as shorter calls, fewer credential asks and lower conversion, are compositional or vanish under adjustment, and we report them as nulls. Beneath the campaigns sit shared assets: the same compliance recording in six campaigns, the same synthetic voice in nine.

A last implication concerns where the detection burden should sit. On AI-generated documents, annotators see artifacts more readily than models yet detect the forgery less often, because the decisive evidence is verifiable rather than visible~\cite{zhang2026receipt}. If synthetic voice on a narrowband line is similarly hard to hear~\cite{mai2023humans}, screening belongs in software rather than the callee's ear. Small on-handset models are plausible candidates, though benchmarking them as call screeners shows scam-side triage trading off sharply against false alarms on legitimate callers~\cite{gan2026callscreenbench}; authenticating the caller at the protocol level is the complementary route~\cite{reaves2017authenticall}.

Three results survive whatever the measurements of §5.5 return, each with a stated limit. A seventh of the calls our persona greeted open with a recording we heard on another call, a floor whose exact level moves with the fingerprint threshold. The identifiable unit of an operation is its opening script and the assets behind it. Its number is disposable, but no more so inside a campaign than in this corpus generally, and the campaign result rests on a clustering choice (§4.7). And a honeypot's seeding history manufactures a trend for anyone not controlling for it, shown here on what amounts to two independent units.

\appendix
\section{Threats to validity}
\begin{figure}[!t]\centering\includegraphics[width=\columnwidth]{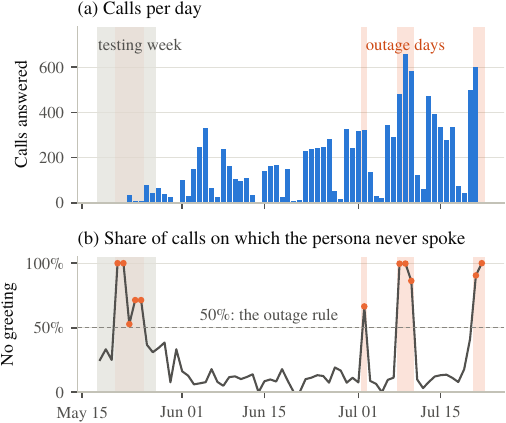}
\caption{\textbf{Every recorded day, and the days our own stack failed.} Calls answered per day (bars) and the share on which the persona never spoke (line). Days on which more than half of the calls got no greeting are shaded and excluded from the decomposition of §4.1 under the rule of §3.1; the gray band on the left is the ten-day pre-corpus testing period.}\label{fig:timeline}\end{figure}

\textbf{The validation study is partial, and its shape bounds what it proves.} The deployed listening pool contains no human-truth negative class, so it measures precision only (§3.4). The gold controls presently contain no known-human audio, so they do not currently catch a listener who answers ``synthetic'' throughout. Agreement between listeners is fair at best (mean pairwise $\kappa$ = +0.267), in line with laboratory findings that people detect speech deepfakes unreliably~\cite{mai2023humans,muller2022perception}, and every clip-level quantity we report inherits that noise.

\textbf{What the judgment export contains.} Every number in §3.4, §4.2 and this appendix is computed from the judgment export that ships with the analysis code. That export holds 3,188 judgments on the ten-second pool, 3,173 of them from eleven named listeners and 15 carrying no name, all of which join to a scored call through the released clip-to-call crosswalk. That is enough to estimate detector precision on the flagged stratum (§4.2).

Two components of a prevalence estimate are missing from it. The unflagged stratum carries only 23 decisive judgments, so the false-negative rate, and therefore corpus prevalence, is unestimated. And the gold controls, 8 judgments in all, contain no known-human audio. Round two holds 521 judgments on 430 clips; its reading is below and is not pooled with the precision estimate.

\textbf{What the listening study does not contain.} The gold set holds no known-human speech over the inbound channel, so a listener who answers ``synthetic'' throughout would score perfectly on it. No clips were labeled by listeners external to the project. And disagreements between listener and detector were not re-labeled against a matched sample of agreements; auditing only disagreements reduces listener noise exclusively in the cells that disagree with the detector and biases measured precision upward, the check the control arm of round two applies.

The interface also gave listeners no definition of either class, no practice clips, and no rule for a recorded notice followed by a live voice. A third label, \emph{recording}, with practice clips belongs to future work (§5.5).

Clips are served in priority tiers that re-serve disagreements, which is why we average per clip rather than per judgment and why we decline to report an agreement coefficient computed within a fixed label-count group. Two of eleven listeners produced judgments with under two seconds of audio consumed on a substantial share of clips; we report listening-duration sensitivity rather than discarding them silently. Finally, the ten-second window is mostly non-speech: a median of 4.1 seconds of speech across all screened clips and 5.8 seconds across the clips listeners judged. Listeners are therefore deciding on less evidence than the window length suggests, a limitation of the stimulus, not of the listeners, and the one round two is designed to remove (§3.4).

\textbf{Round two.} The twenty-second round holds 521 judgments, all but 19 from two of the round-one listeners: 392 on the 330 contested clips and 129 on the 100 control clips, a median of one judgment per clip. Both arms moved toward ``synthetic.'' The mean per-clip synthetic fraction rose from 0.56 to 0.79 on contested clips and from 0.39 to 0.56 on controls. The rise holds in each listener separately: 65\% and 90\% ``synthetic'' on contested clips against 39\% and 79\% on controls. The contested-minus-control difference in that shift is +0.06 (clip-bootstrap 95\% CI −0.05 to +0.16). Among the 27 clips with at least two decisive judgments in both rounds, three majorities flipped from human to synthetic and none the other way (McNemar $p = 0.25$).

With one judgment per clip, polarization (how far a clip's synthetic fraction sits from an even split) is 1 by construction and cannot yet speak to whether disagreement was resolved. The reading is a criterion shift, in which more speech moves both listeners toward ``synthetic.'' The contested-minus-control difference is indistinguishable from zero at this sample size, and there is no evidence yet that disagreement on the contested clips was resolved. Answering that question would need at least three judgments per clip, which the round did not reach, and its judgments are not pooled with the precision estimate above. Round two also bears on the headline. If the same clips move from 0.56 to 0.79 when the window grows from ten to twenty seconds, the 54.4\% of §4.2 is partly a property of the ten-second stimulus. It should then be read as precision conditional on the window the detector scored.

\textbf{The instrument.} Published detectors collapse on in-the-wild audio. AASIST~\cite{jung2022aasist} falls from an in-domain AUC (area under the receiver operating characteristic curve) of 1.00 to 0.43, no commercial model reaches 90\% accuracy~\cite{chen2025deepfakeeval}, and the gap is almost entirely domain shift rather than attack hardness~\cite{muller2024domainshift,muller2022generalize}. Narrowband telephony is the harder end of that problem~\cite{asvspoof2021}, and codec compression costs a mean 5.30 points of equal-error rate~\cite{addc2025}. Even a detector trained on presented and augmented audio still misses 23.7\% of deepfakes at a 1\% false-alarm rate when the audio is played through a loudspeaker into a real call. When the audio is injected into the call directly, it misses 11.8\%~\cite{delgado2025playback}.

The failure is not peculiar to speech: image deepfake detectors on real-service faceswaps likewise fall to near chance~\cite{ren2025realitydeepfake}, and the benchmarks such detectors are trained on have structural limits of their own~\cite{layton2024sok}. We cannot disentangle how much of any score reflects the voice versus the line it traveled over, which carries measurable acoustic signatures of its own~\cite{balasubramaniyan2010pindrop}. The line's share is bounded from below: 13.6\% of the 2,756 identical-waveform pairs in this corpus fall on opposite sides of the threshold (§4.2). Every scored window is also post-processed audio. The telephony provider applies noise suppression to the inbound leg before it reaches our recorder, so the detector never heard the raw line. Whatever suppression does to synthesis artifacts is done to every call alike.

\textbf{A shared-instrument null.} The detector API also returns an accent tag. Over all 1,816 AI-labeled calls, replays included (§4.8 gives the fresh-audio split), 92.2\% carry a generic ``American'' tag against 71.7\% of human-labeled calls, a direction holding across four views (AI 72--96\%, human 53--85\%). We report this as a property of the API rather than of the callers. Tag and score come from the same system on the same clip, and conditioning harder on detector confidence \emph{widens} the gap, which is the signature of a shared-feature confound, not of an independent instrument. Establishing an acoustic monoculture would need a second accent labeler.

\textbf{Temporal coverage.} We score ten seconds of caller speech from the onset of the first utterance, which is 7.3\% of the median 138-second call. Every label in this paper is therefore a property of a call's \emph{opening}, not of the whole conversation. A caller that is synthetic for ten seconds and hands to a live agent at thirty is counted as synthetic, and we cannot detect a voice change later in the call. Scoring only the opening does not undermine the comparisons, which apply the same rule everywhere, but it bounds what they are comparisons \emph{of}. It also means our prevalence is best read as the share of calls that \emph{open} with a non-live voice. Scoring several windows per call would settle how often the voice changes (§5.5).

\textbf{Construct.} The detector labels what it labels; the decomposition of §4.1 is what the paper claims. Two of its rows rest on audio evidence alone: a recording heard on another call, and fresh audio the detector labels synthetic. One rests on an inference, that a caller silent for a median of 16 seconds after our greeting is machine-placed. That inference rests on the persona speaking first and never re-prompting (§3.1), and on the premise that a person who wanted the call says something. A person who dialed by mistake and listened in silence is counted as a machine.

The replay row still does not say whether the replayed recording was made by a person or rendered once by a synthesizer. The detector's label on those calls (80.7\% synthetic) is the only evidence, and the detector is the instrument under test. Readers should treat 13.1\% as synthesis prevalence on fresh audio, 13.8\% as automated playback of unknown origin, and 26.9\% as the machine-voiced share both rows support.

\textbf{Outage days and dead air.} On eleven of 66 days the honeypot answered but never spoke, and on the other days 12.6\% of calls got no greeting either (Figure~\ref{fig:timeline}). We class a day as an outage when more than half of its calls carry no persona speech, and we exclude those 2,711 calls from the decomposition. A caller's behavior toward dead air is not the behavior toward a greeting that Table~\ref{tab:decomp} describes. The outage days fall in the ten-day pre-corpus testing period and on six days in July, when the bait numbers were oldest, so their exclusion removes more old-number than new-number traffic. The age effect of §4.4 is computed on the detector-scored corpus and does not depend on that exclusion.

\textbf{Replay detection has a hole where the smart operators are.} The replay census finds a recording only when it is played to us at least twice. A call center that dials each number once and never again, which is what a careful operator does, leaves one waveform and no cluster. Its recordings sit in the ``fresh audio'' rows of Table~\ref{tab:decomp} wearing whatever label the detector gives them. Two signals in §4.8 need no second call, turn-taking and within-call self-replay, and they separate the classes, but neither is yet a classifier.

The 13.8\% is therefore a floor on playback at the 0.70 criterion, and 9.2\% at the 0.85 criterion (§4.1). The split between playback and synthesis inside the detector's 29.3\% is the one number in this paper we expect to move most when a better instrument arrives. A word-aligned comparison of the words two calls share would extend the census to templates with a name spliced in; we have not run it.

\textbf{Persona bleed-through.} The caller track could carry an echo of our own persona through the far end's device or network. We tested it directly: each scored window was slid along the persona track of its own call and along another call's persona track, taking the best normalized cross-correlation in each. On the listening pool, 9 clips exceeded the null's 99.9th percentile against 2.4 expected by chance. Over all 8,097 recorded calls with both tracks, 63 exceeded the 99th percentile against 81 expected and 6 the 99.9th against 8. A speaker-embedding check found 8 of 1,733 clips with a persona reference whose voice is indistinguishable from that persona's; they are listed for listening. The scored audio does not carry our voice.

\textbf{Threshold and aggregation.} The threshold was chosen on the same data it is applied to and is not at the distribution's trough (§3.3), and the per-utterance maximum is a logical-OR rule whose false-positive rate grows with utterance count. Together these move the estimate across 21.8--39.6\%.

\textbf{Population.} The numbers were actively seeded into lead-generation funnels in a narrow vertical rather than passively observed. The sample is therefore the downstream buyers of one synthetic lead identity, not the calling ecosystem, and only 15\% of classified calls meet our fraud criterion. Calls reaching an answering honeypot are not the population a campaign originates. Nor is the traffic conditioned much on the caller failing to notice us. The ending labeler marks the persona as detected on 320 of the 6,192 scored calls (5.2\%), close to the 5\% recognition rate reported for Lenny~\cite{sahin2017lenny}, and those calls are less often AI-labeled (21.6\%) than the rest (29.8\%). The two-turn corpus filter removes silent connections and one-line hang-ups rather than prerecorded blasts (§4.1); the decomposition counts them, the detector-labeled rates of §4.2 onward do not.

The seeding schedule is itself part of the sample: Figure~\ref{fig:exposure} plots synthetic share against each receiving number's exposure age, the confound §4.4 describes. Our class labels are our own classifier's. The companion descriptor~\cite{scamai2026corpus} reports its holistic pass agreeing with human reviewers on 75\% of binary scam-versus-not judgments, against 67\% for the strict label, figures the public analysis of this corpus~\cite{traister2026anatomy} relays pending the descriptor's release. Roughly a quarter of the assignments underlying Table~\ref{tab:classes} are therefore wrong, and we do not propagate that into its intervals.

\textbf{Inference.} This paper reports roughly fifty comparisons without family-wise correction and the specification quoted varies by result, so the analysis should be read as exploratory. For orientation, under a Bonferroni factor of 50 the clustered robocall-opening enrichment of §4.6 survives marginally ($p = 0.049$). The stratified duration and turn-count comparisons of §4.6 do not survive at all, and §4.5 reports a count within the flagged population and attaches no test. Clustering on originating number is applied where reported but not uniformly, and no analysis clusters on campaign, which §4.7 shows is the stronger dependence. A related unit-of-analysis caveat attaches to every per-number quantity: the detector labels calls, not numbers, and 187 numbers carry calls of both labels.

\section{Ethics}

The corpus was generated by submitting fabricated contact details into third-party lead-generation forms, some run by legitimate businesses, which injects false records into their databases and consumes their outbound capacity. We judge the harm small relative to the measurement, since the volume is low, the records inert, and the funnels resell contact data into precisely the campaigns under study. It is nonetheless a real cost imposed on third parties.

The personas are designed to deceive the callers who reach them. The justification is the scambaiting literature's, adapted to a corpus that is mostly telemarketing rather than fraud (only 15\% of classified calls meet our fraud criterion, Table~\ref{tab:classes}). The personas answer only calls placed to numbers we own, the persona and every detail it offers are fictitious, and the caller's cost is the time spent on a lead that was never real. We do not rest the justification on the caller's intent, which we cannot verify. The deception is real and deliberate, and justified by that asymmetry rather than by any claim that no deception occurs.

The audio records a real, non-consenting individual, and what that caller was attempting does not extinguish every interest they retain in it. Listeners hear a ten-second opening (round one) or roughly twenty seconds of speech with silence removed (round two, §3.4), never a full conversation. Round-one clips passed through an automated pass that mutes full names, phone and account numbers, street addresses, ZIP codes, email addresses, URLs and dates of birth before being served. That pass has been verified by a level check on each masked span rather than by listening back. Round-two clips were served without that pass, to listeners recruited from within the project.

De-identification is applied at the point of exit. Opener text used for clustering has personal first names removed (§4.7), and every released table carries caller numbers as hashes. Any clip or transcript excerpt released beyond the study passes a listening-verified screen for identifiable personal information before release. We do not promise a specific release form. We do commit unconditionally to releasing the analysis code, per-call derived features and labels with caller numbers hashed, the annotation interface, and the clip-extraction parameters, none of which requires releasing audio. The banner and answering-beside-the-point counts derive from the honeypot's message log, which is released as per-call derived values rather than verbatim text.

Finally, the dual-use risk. The FCC's 2024 ruling placed AI-generated voices under the TCPA with no published baseline behind it; the policy already exists and is enforced in the absence of this measurement. Withholding it would not undo the policy, only leave regulators acting on assumption.

\label{endbody}
\section*{Data and Code Availability}
Supporting code and the non-restricted resources for this work are available from the Scam.AI
Research portal, \url{https://scam.ai/research}~\cite{scamai_research}, with document- and
PDF-forgery resources maintained at Trustable, \url{https://trustable.ai}~\cite{trustable_ai}.
As noted in the paper, the raw audio recordings used in this study are not publicly released;
requests and further details are handled through the Scam.AI Research portal.

\bibliographystyle{unsrt}\bibliography{refs}
\end{document}